\documentclass[11pt]{article}
\usepackage{epsfig}
\usepackage{amssymb}
\usepackage{amsmath}
\usepackage{amsfonts}
\newcommand{\R}{\mathbb{R}}
\newcommand{\C}{\mathbb{C}}

\newcommand{\cC}{\cal C}

\newcommand{\cH}{\cal H}

\newcommand{\rp}{{\rm p}}

\newcommand{\fcc}{f^{(\vec\zeta_\epsilon,\epsilon)}}
\newcommand{\fc}{f_{_{\rm coh}}}
\newcommand{\psiac}{\psi_a^{(\epsilon,\vec\zeta_\epsilon)}}

\newcommand{\vm}{\texttt{\textit{m}}}
\newcommand{\be}{\begin{equation}}
\newcommand{\ee}{\end{equation}}
\newcommand{\bea}{\begin{eqnarray}}
\newcommand{\eea}{\end{eqnarray}}
\newcommand{\nn}{\nonumber}
\newcommand{\kt}{\rangle}
\newcommand{\br}{\langle}

\newcommand{\cum}{\mbox{\scriptsize${\cal M}$}}
\newcommand{\ed}{\end{document}}

\newcommand{\aeq}{&\!\!\!\!=\!\!\!\!&}
\newcommand{\adef}{&\!\!\!\!:=\!\!\!\!&}
\newcommand{\romx}{\vec{\rm x}}

\begin{document}

\title{Quantum Mechanics of Klein-Gordon Fields II:\\
Relativistic Coherent States}
\author{A.~Mostafazadeh\thanks{Corresponding author, E-mail address:
amostafazadeh@ku.edu.tr}~and  F.~Zamani\thanks{E-mail address:
zamani@iasbs.ac.ir} \\ \\
$^*$~Department of Mathematics, Ko\c{c} University,
Rumelifeneri Yolu, \\ 34450 Sariyer, Istanbul, Turkey \\
$ ^\dagger$~Department of Physics, Institute for Advanced Studies
in Basic \\ Sciences, 45195-1159 Zanjan, Iran}
\date{ }
\maketitle

\begin{abstract}
We use the formulation of the quantum mechanics of first quantized
Klein-Gordon fields given in the first of this series of papers to
study relativistic coherent states. In particular, we offer an
explicit construction of coherent states for both charged and
neutral (real) free Klein-Gordon fields as well as for charged
fields interacting with a constant magnetic field. Our
construction is free from the problems associated with
charge-superselection rule that complicated the previous studies.
We compute various physical quantities associated with our
coherent states and present a detailed investigation of their
classical (nonquantum) and nonrelativistic limits.
\end{abstract}

\section{Introduction}

The study of the relationship between classical and quantum
mechanics (QM) has been among the most important issues of modern
theoretical physics. A related subject of great significance that
has been the focus of attention since the very early days of QM is
that of coherent states. Introduced by Schr\"odinger
\cite{schrodinger} as early as in 1926 and systematically
formulated and developed in the capable hands of Glauber
\cite{glauber}, Sudarshan \cite{sudarshan}, and Klauder
\cite{klauder1} in the 1960s, coherent states have found
widespread applications in various branches of physics, extending
from particle physics to quantum optics
\cite{klauder-b,perelomov-b,zhang,feng} and more recently quantum
computation \cite{q-computation}.

The conceptual and practical significance of coherent states have
naturally motivated the introduction and study of their
generalizations. Probably the most notable of these has been the
development of coherent states associated with general Lie
algebras \cite{klauder1,perelomov1,gilmore,perelomov-b,zhang}.
Another basic problem has been to explore the relativistic
coherent states. The purpose of the present paper is to employ the
formulation of the QM of Klein-Gordon (KG) fields developed in
\cite{p59}, to provide an explicit construction of relativistic
coherent states for first-quantized KG fields
\cite{malkin,bagrov,lev-colab,haghighat}.\footnote{For a
discussion of coherent states for second-quantized scalar fields
see \cite{botke,debajyoti,skagerstam}. Some other related
publications are \cite{aldaya,tang,field}.}

Our basic point of departure, besides the general formalism
developed in \cite{p59}, is Glauber's description of
nonrelativistic coherent states \cite{glauber}, according to which
one may use any one of the following equivalent definitions: 1)
The coherent state vectors $|\zeta\kt$ are eigenvectors of the
(harmonic-oscillator) annihilation operator $a$, $a|\zeta\kt =
\zeta|\zeta\kt$; 2) $|\zeta\kt$ can be obtained by applying the
Glauber's displacement operator $D(\zeta):=\exp(\zeta
a^\dag-\zeta^* a)$ on the vacuum state vector $|0\kt$, $|\zeta\kt
= D(\zeta)|0\kt$; 3) $|\zeta\kt$ determines a quantum state with a
minimum-uncertainty relationship, $\Delta p~\Delta q = \hbar/2$,
where $q, p$ are the canonically conjugate coordinate and momentum
operators, $\Delta O:=\sqrt{\br\zeta|O^2|\zeta\kt-
\br\zeta|O|\zeta\kt^2}$ for an observable $O$, and $|\zeta\kt$ is
assumed to be normalized.

A key issue in the construction of coherent states is the
identification of an appropriate canonically conjugate pair of
coordinate and momentum operators. For a nonrelativistic particle
having $\R^n$ as its configuration space, these are usually the
position and momentum operators in a Cartesian coordinate system.
But in general one may make other choices. A typical example is
the choice made in Re.~\cite{malkin} where the authors use the
symmetries of the problem to select a pair of annihilation
operators that do not correspond to the usual choice of coordinate
and momentum operators. The problem of determining coordinate and
momentum operators is at the heart of much of the difficulties
encountered in constructing a physically viable candidate for
relativistic coherent states. This is mainly because the
identification of an appropriate relativistic position operator
has been a nontrivial task
\cite{newton-wigner,other-position,fesh-vil,p57,p59}.

The first thorough investigation of relativistic coherent states
for first quantized KG fields is due to Bagrov, Buchbinder, and
Gitman \cite{bagrov}. These authors express the KG equation in a
certain null plane coordinate system as an equation which is first
order in one of the coordinates and employ the idea of linear
dynamical invariants of Malkin and Man'ko \cite{malkin} to obtain
a set of relativistic coherent states. These coherent states have
certain appealing properties, but the corresponding coordinate
operator does not coincide with any of the known relativistic
position operators \cite{newton-wigner,other-position}.

In a more recent paper \cite{lev-colab}, Lev, Semenov, Uslenko,
and Klauder construct a set of coherent states for KG fields
interacting with a static magnetic field. Their analysis uses the
position operator $q$ of \cite{fesh-vil,greiner} that does not
respect the charge superselection rule \cite{wick}, i.e., its
action mixes the negative- and positive-energy KG fields. To
remedy this problem, Lev et al choose to construct the
relativistic coherent states, for free KG fields, using the ``even
part'' (the part that respects the charge superselection rule) of
this position operator. This corresponds to the celebrated
Newton-Wigner position operator \cite{newton-wigner}. The
identification of $q$ with an observable causes certain
peculiarity such as the possibility of a negative value for
$\br\zeta|q^2|\zeta\kt$ and a subsequent violation of the minimum
uncertainty relation.

The approach of \cite{lev-colab} for treating coherent states for
a KG field interacting with a constant magnetic field is based on
the idea of nonlinear coherent states \cite{matos}. It involves
the introduction of a deformed annihilation operator whose even
part is used to construct a set of relativistic coherent states.
This leads to some complications as the even part of the deformed
annihilation and creation operators do not satisfy the usual
commutation relations. Furthermore, this construction which relies
on decoupling the translational and rotational degrees of freedom
encounters the difficulty that the even part of the annihilation
operators associated with these two degrees of freedom do not
commute. This in turn necessitates the consideration of separate
sets of coherent states corresponding to translational and
rotational motions. These also display peculiar behaviors
\cite{lev-colab}.

Refs.~\cite{p57,p59} report on the construction of a set of
relativistic position and momentum operators that respect the
charge superselection rule. The eigenstates of the corresponding
annihilation operator have a definite charge. Hence, they
represent a set of coherent states that are free from the
difficulties encountered in the above-mentioned studies. The
purpose of the present paper is to perform a comprehensive
investigation of the properties of these coherent states.

As we show in \cite{p59}, we can apply our formulation for scalar
fields interacting with an arbitrary stationary magnetic field by
enforcing the minimal-coupling prescription,
$\vec\nabla\rightarrow\vec\nabla-ie\vec A(\vec x)$, in our
treatment of free fields. In the presence of a nonstationary
magnetic or a nonzero electric field, we are obliged to employ the
nonstationary QM outlined in \cite{ap} and deal with the fact that
the time-evolution is necessarily non-unitary. Therefore, in this
paper we only consider, besides the free scalar fields, the
charged scalar fields interacting with a constant homogeneous
magnetic field.

The organization of the article is as follows. In Section 2 we
outline a general construction for coherent states of a charged
relativistic particle. In Section 3, we focus our attention on the
coherent states of a free relativistic particle and examine their
physical properties and classical and nonrelativistic limits. In
Section 4, we study the coherent states of a neutral scalar
particle. In Section 5, we consider the consequences of coupling a
complex scalar field to a constant homogeneous magnetic field.
Finally, in Section 6 we present our concluding remarks.

Throughout this paper we will occasionally refer to
Ref.~\cite{p59} as paper I and use the label (I-n) to denote
Eq.~(n) of paper I. Furthermore, we recall from paper I that the
free KG equation may be expressed as
    \be
    \ddot\psi(x^0)+D\psi(x^0)=0,
    \label{kg}
    \ee
where $\psi(x^0):\R^3\to\C$ is defined by $\psi(x^0)(\vec
x):=\psi(x^0,\vec x)$ for all $x=(x^0,\vec x)\in\R^4$, an overdot
stands for an $x^0$-derivative, $D:L^2(\R^3)\to L^2(\R^3)$ is the
operator defined by $(D\phi)(\vec x):=(-\nabla^2+\cum^2) \phi(\vec
x)$ for all $\phi\in L^2(\R^3)$, and $\cum:=mc/\hbar$. As we
mentioned earlier we may account for the interaction with a
stationary magnetic field by letting
$\vec\nabla\to\vec\nabla-ie\vec A(\vec x)$ in the preceding
formula for $D$, where $\vec A$ is the vector potential.

\section{General Constructions for a Charged Scalar Field}

Following paper I, Let ${\cal H}_a$ be the Hilbert space obtained
by endowing the vector space ${\cal V}$ of solutions $\psi$ of the
free KG equation (\ref{kg}), with the inner product
    \bea
    (\psi_1,\psi_2)_a&=&\frac{\kappa}{2\cum}\,
    \left\{\br\psi_1(x^0)|D^{1/2}\psi_2(x^0)\kt
    +\br\dot\psi_1(x^0)|D^{-1/2}\dot\psi_2(x^0)\kt+\right.\nn\\
    &&\hspace{1.5cm}\left. ia\left[\br\psi_1(x^0)|
    \dot\psi_2(x^0)\kt-\br\dot\psi_1(x^0)|
    \psi_2(x^0)\kt\right]\right\},
    \label{gen-i}
    \eea
where $a\in(-1,1)$, $\kappa\in\R^+$, and $\psi_1,\psi_2\in{\cal
V}$ are arbitrary. As shown in \cite{p57} (See Eq.~(I-42)), the
Hilbert space ${\cal H}_a$ may be mapped onto the Hilbert space
${\cal H}':=L^2(\mathbb{R}^3)\oplus L^2(\mathbb{R}^3)$ of the
two-component Foldy representation by a unitary transformation
$U_a:{\cal H}_a\to {\cal H}'$ of the form
    \bea
        U_a\psi \adef\frac{1}{2}\sqrt{\frac{\kappa}{\cum}}
            \left(\begin{array}{c}
            \sqrt{1+a}~[D^{1/4}\psi(x^0_0)+iD^{-1/4}\dot\psi(x^0_0)]\\
            \sqrt{1-a}~[D^{1/4}\psi(x^0_0)-iD^{-1/4}\dot\psi(x^0_0)]
            \end{array}\right)\nn\\
        \aeq\frac{1}{2}\sqrt{\frac{\kappa}{\cum}}\,D^{1/4}
            \left(\begin{array}{c}
            \sqrt{1+a}~[\psi(x^0_0)+\psi_c(x^0_0)]\\
            \sqrt{1-a}~[\psi(x^0_0)-\psi_c(x^0_0)]
            \end{array}\right),
        \label{U-a}
    \eea
where $x^0_0\in\mathbb{R}$ is a fixed initial value for $x^0=c t$
and $\psi_c(x^0):=iD^{-1/2}\dot\psi(x^0)$. The inverse of $U_a$ is
given by
    \be
    \left[U_a^{-1} \xi\right](x^0)
    =\sqrt{\frac{\cum}{\kappa}}\;D^{-1/4}
       \left[e^{-i(x^0-x^0_0)D^{1/2}} \frac{\xi^1}{\sqrt{1+a}}
       +e^{i(x^0-x^0_0)D^{1/2}} \frac{\xi^2}{\sqrt{1-a}} \right],
    \label{U-a-inv}
    \ee
where $\xi=\mbox{\tiny $\left(\begin{array}{c} \xi^1\\
\xi^2\end{array}\right)$}\in{\cal H}'$ and $x^0\in\mathbb{R}$ are
arbitrary.

Denoting the $2\times 2$ identity matrix by $\sigma_0$, we can
respectively express the position and momentum operators acting in
${\cal H}'$ as
    \be
    \vec X':=\vec{\rm x}\otimes\sigma_0,~~~~~~~~~~
    \vec P':=\vec{\rm p}\otimes\sigma_0,
    \label{xp-prime}
    \ee
where $\vec{\rm x}$ and $\vec{\rm p}$ are the ordinary position
and momentum operator acting in $L^2(\R^3)$. Next, we define the
position ($\vec X_a$) and momentum ($\vec P_a$) operators and the
associated annihilation operator ($\vec A_a$) acting in the
Hilbert space ${\cal H}_a$ of the one-component KG fields by
    \bea
    \vec X_a&:=& U_a^{-1}\vec X'\,U_a,~~~~~~~
    \vec P_a:= U_a^{-1}\vec P'\,U_a,
    \label{hxp}\\
    \vec A_a&:=&\sqrt{\frac{k}{2\hbar}}\,\left(\vec X_a+ik^{-1}\vec P_a\right),
    \label{ann}
    \eea
where $k=\vm\,\omega\in\mathbb{R}$ is the characteristic
oscillator constant having the dimension of mass per time,
\cite{p57}. We then identify the coherent states with the
eigenstates of $\vec A_a$.

The charge-grading operator ${\cal C}'$ and the charge
operator\footnote{See \cite{greiner} page 75.} ${\cal Q}'$ acting
in $\cH'$ are given by
    \be
    {\cal C}'=\sigma_3=\left(\begin{array}{cc}
    1 & 0 \\
    0 & -1\end{array}\right), ~~~~~~~~~~~~{\cal Q}': = e \sigma_3 ,
    \label{cg}
    \ee
where $e$ is the unit charge \cite{p57}. In view of
(\ref{U-a-inv}) and ${\cal C}=U_a^{-1}{\cal C}'U_a$, it is
straightforward to calculate the charge-grading and charge
operator acting in ${\cal H}_a$. The result is \cite{p57, p59}
    \be
    {\cal C} = i D^{-1/2}\frac{\partial}{\partial x^0}, ~~~~~~~~~~~~
    {\cal Q} := e\, {\cal C} = i e D^{-1/2}\frac{\partial}{\partial x^0}.
    \label{hcg}
    \ee
As seen from these relations, ${\cal C}$ and ${\cal Q}$ do not
depend on the parameter $a$. Furthermore, because ${\cal C}',{\cal
Q}':{\cal H}'\to{\cal H}'$ are Hermitian and $U_a:{\cal H}_a\to
{\cal H}'$ is unitary, ${\cal C},{\cal Q}:{\cal H}_a\to{\cal H}_a$
are Hermitian as well, i.e., they are physical observables.

Because both $\vec X_a$ and $\vec P_a$ commute with the
charge-grading operator ${\cal C}$, so does the annihilation
operator $\vec A_a$. This means that they respect the charge
superselection rule \cite{p57, p59} and that the corresponding
coherent states may be taken to have a definite charge. We will
identify the latter with the common eigenstates of $\vec A_a$ and
$\cC$. The associated state vectors
$\psi_a^{(\epsilon,\vec\zeta_\epsilon)}$ satisfy
    \be
    \vec A_a\: \psi_a^{(\epsilon,\vec\zeta_\epsilon)} =
    \vec\zeta_\epsilon\: \psi_a^{(\epsilon,\vec\zeta_\epsilon)},
    ~~~~~~~~~~~~
    {\cal C}\: \psi_a^{(\epsilon,\vec\zeta_\epsilon)} =
    \epsilon \: \psi_a^{(\epsilon,\vec\zeta_\epsilon)},
    \label{hcoh}
    \ee
where $\vec\zeta_\epsilon\in\mathbb{C}^3$ and $\epsilon=\pm$. By
construction, the coherent state vectors
$\psi_a^{(\epsilon,\vec\zeta_\epsilon)}$ are free from the
peculiarities associated with the nontrivial charge structure of
the conventional coherent states \cite{botke, debajyoti,
skagerstam, malkin, bagrov, lev-colab}. They are also solutions of
the KG equation (\ref{kg}). We will refer to them as {\em coherent
Klein-Gordon fields}.

Because the position operator $\vec X_a$ has a complicated form
\cite{p57,p59}, Eqs.~(\ref{hcoh}) do not offer a practical method
of computing $\psi_a^{(\epsilon,\vec\zeta_\epsilon)}$. A more
convenient method is to construct the corresponding coherent state
vectors in the Foldy Representation, namely
    \be
    |\vec\zeta_\epsilon,\epsilon\kt:=
    U_a\psi_a^{(\epsilon,\vec\zeta_\epsilon)}.
    \label{z-epsilon}
    \ee

\subsection{Coherent States in Foldy Representation}

In the Foldy representation, where the Hilbert space is ${\cal
H}'=L^2(\mathbb{R}^3)\oplus L^2(\mathbb{R}^3)$, the annihilation
operator has the form
    \be
    \vec A':=U_a\,\vec A_a\,U_a^{-1}=\sqrt{\frac{k}{2\hbar}}\,
    \left(\vec X'+ik^{-1}\vec P'\right).
    \label{fann}
    \ee
In view of (\ref{hxp}) -- (\ref{cg}), (\ref{z-epsilon}) and
(\ref{fann}),
    \be
    \vec A'|\vec \zeta_\epsilon,\epsilon\kt=
    \vec \zeta_\epsilon|\vec \zeta_\epsilon,\epsilon\kt,
    ~~~~~~~~~
    {\cal C}'|\vec \zeta_\epsilon,\epsilon\kt=
    \epsilon|\vec \zeta_\epsilon,\epsilon\kt,
    ~~~~~~~~~|\vec \zeta_\epsilon,\epsilon\kt=
    |\vec \zeta_\epsilon\kt\otimes e_\epsilon,
    \label{zet}
    \ee
where $e_+:=\mbox{\tiny $\left(\begin{array}{c} 1\\
0\end{array}\right)$}$, $e_-:=\mbox{\tiny $\left(\begin{array}{c} 0\\
1\end{array}\right)$}$, and $|\vec \zeta_\epsilon\kt$ are the
ordinary coherent state vectors satisfying
    \be
    \vec a|\vec \zeta_\epsilon\kt=\vec \zeta_\epsilon |\vec
    \zeta_\epsilon\kt,~~~~{\rm with}~~~~
    \vec a:=\sqrt{\frac{k}{2\hbar}}\,
    \left(\vec{\rm x}+ik^{-1}\vec{\rm p}\right).
    \label{nonrel-coh}
    \ee

Next, we use the (complete and orthonormal) position basis vectors
$\xi_{\epsilon,\vec x}:=|\vec x\kt\otimes e_\epsilon$ of ${\cal
H}'$ to introduce the coherent wave function
$f^{(\vec\zeta_\epsilon,\epsilon)}$ as
    \be
    \fcc(\epsilon',\vec x):=
    \br\xi_{\epsilon',\vec x} | \vec \zeta_\epsilon,\epsilon\kt=
    \br\vec x|\vec \zeta_\epsilon\kt\,\delta_{\epsilon',\epsilon}=
    \fc(\epsilon,\vec x)\,\delta_{\epsilon',\epsilon},
    \label{fxz}
    \ee
where $\epsilon,\epsilon'=\pm$ and
    \be
    \fc(\epsilon,\vec x):=\br\vec x|\vec \zeta_\epsilon\kt=
    \fcc(\epsilon,\vec x).
    \label{fc-define}
    \ee
As seen from this relation $\fc(\epsilon,\vec x)$ is the usual
coherent state wave function of the nonrelativistic QM which
according to (\ref{nonrel-coh}) satisfies
    \be
    \sqrt{\frac{k}{2\hbar}}\,
    \left(\vec x+k^{-1}\vec\nabla\right)\fc(\epsilon,\vec x)=
    \vec\zeta_\epsilon \fc(\epsilon,\vec x).
    \label{eket2}
    \ee
In view of (\ref{fxz}) and the completeness of the position kets
$|\vec x\kt$ (alternatively $\xi_{\epsilon,\vec x}$),
    \be
    |\vec \zeta_\epsilon,\epsilon\kt=\int_{\mathbb{R}^3} d^3 x \,
    |\vec x\kt\br\vec x|\vec \zeta_\epsilon\kt\otimes e_\epsilon
    =\int_{\mathbb{R}^3} d^3 x \,\fc(\epsilon,\vec x)\,\xi_{\epsilon,
    \vec x}.
   \label{ss}
   \ee

Similarly, we can employ the momentum basis vectors
$\xi_{\epsilon,\vec p}$ of ${\cal H}'$, that fulfil
    \be
    \xi_{\epsilon,\vec p}:=|\vec p\kt\otimes e_\epsilon,~~~~~~~
    \br \xi_{\epsilon,\vec p},\xi_{\epsilon',\vec p'}\kt=
    \delta_{\epsilon,\epsilon'}\delta^3(\vec p-\vec p'),~~~~~~~
    \sum_{\epsilon=\pm}\int_{\mathbb{R}^3}d^3p\:|\xi_{\epsilon,\vec p}\kt
    \br\xi_{\epsilon,\vec p}|=\sigma_0,
    \label{mom-bas}
    \ee
to obtain the momentum representation of coherent states, namely
$|\vec\zeta_\epsilon,\epsilon\kt =\int_{\mathbb{R}^3} d^3 p \,
\fc(\epsilon,\vec p)\,\xi_{\epsilon,\vec p}$, where
   \be
    \fc(\epsilon,\vec p):=\br\vec p|\vec\zeta_\epsilon\kt =
    \br\xi_{\epsilon,\vec p} | \vec \zeta_\epsilon,\epsilon\kt
    = \frac{1}{(2\pi\hbar)^{3/2}}\int d^3 x\:
    e^{-i\frac{\vec p}{\hbar}\cdot\vec x}\:\fc(\epsilon,\vec x).
   \label{fmom}
   \ee

As it is well-known, $|\vec \zeta_\epsilon\kt$ are normalizable,
non-orthogonal, and overcomplete \cite{glauber,zhang}. Moreover
they yield a resolution of the identity which is not unique
\cite{glauber}. A common choice for the latter is
   \be
    \int |\vec\zeta_\epsilon\kt
    \frac{d^2\zeta_\epsilon}{\pi} \br\vec\zeta_\epsilon| = 1 ,
    \label{resol2}
   \ee
where $d^2\zeta_\epsilon = d \Re(\zeta_\epsilon)\; d
\Im(\zeta_\epsilon)$.\footnote{Here and in what follows $\Re$ and
$\Im$ respectively means `real' and `imaginary part of'.}

Two-component coherent state vectors $|\vec\zeta_\epsilon,\epsilon\kt=|\vec
\zeta_\epsilon\kt\otimes e_\epsilon$ share the properties of
$|\vec \zeta_\epsilon\kt$; with an appropriate normalization
constant they satisfy
   \be
   \br\vec\zeta_\epsilon,\epsilon|\vec\zeta'_{\epsilon'},\epsilon'\kt=
   \delta_{\epsilon \epsilon'}\,
   \exp \left[\vec\zeta_\epsilon^*\cdot\vec\zeta'_\epsilon -
   \frac{1}{2}( |\vec\zeta_\epsilon|^2 + |\vec\zeta'_\epsilon|^2 )
   \right]\,,~~~~~~
   \sum_{\epsilon=\pm}\int |\vec\zeta_\epsilon,\epsilon\kt\,
   \frac{d^2\zeta_\epsilon}{\pi}\,\br\vec\zeta_\epsilon,\epsilon|
   =\sigma_0 \,.
   \label{compl}
   \ee
Using these properties, we can express any two-component vector
$\Psi\in{\cal H}'$ in the coherent state basis $\{|\vec
\zeta_\epsilon,\epsilon\kt\}$ as
  \be
  \Psi = \sum_{\epsilon=\pm}\int |\vec\zeta_\epsilon,\epsilon\kt\,
  \frac{d^2\zeta_\epsilon}{\pi}\,\br\vec\zeta_\epsilon,\epsilon|\Psi\kt =
  \sum_{\epsilon=\pm}\int \frac{d^2\zeta_\epsilon}{\pi}\,
  g(\vec\zeta_\epsilon,\epsilon)\,|\vec\zeta_\epsilon,\epsilon\kt\,,
  \label{psizeta}
  \ee
where $g(\vec\zeta_\epsilon,\epsilon) :=
\br\vec\zeta_\epsilon,\epsilon|\Psi\kt$ is the wave function
associated with $\Psi$ in its coherent state representation.

Next, we wish to consider the dynamical aspects of the theory in
the coherent state representation. Recall that in the Foldy
representation, the dynamics is generated by the Schr\"odinger
equation for the Hamiltonian: $H' = \hbar\sqrt{D}\sigma_3 =
\sqrt{\vec{\rm p}^2+m^2c^2}\,\sigma_3$, where
$\sigma_3:=\mbox{\tiny $\left(\begin{array}{cc}
    1 & 0 \\
    0 & -1\end{array}\right)$}$.
The time-evolution of an initial coherent state vector $|\vec
\zeta_\epsilon,\epsilon;x^0_0\kt:=|\vec \zeta_\epsilon,\epsilon\kt$ is given by
    \be
    |\vec\zeta_\epsilon,\epsilon;x^0\kt=
    e^{-i\epsilon(x^0-x^0_0)D^{1/2}}
    |\vec\zeta_\epsilon,\epsilon;x^0_0\kt
    =\int_{\mathbb{R}^3}d^3p\:
    e^{-i\epsilon(\frac{x^0-x^0_0}{\hbar})\sqrt{\vec p^2+m^2c^2}}\:
    \fc(\epsilon,\vec p;x^0_0)\:|\xi_{\epsilon, \vec p}\kt,
    \label{zetatcoh}
    \ee
where we have used (\ref{zet}) and (\ref{mom-bas}). The coherent
wave function $\fc(\epsilon,\vec x;x^0):=\br\xi_{\epsilon,\vec x}
    |\vec\zeta_\epsilon,\epsilon;x^0\kt$ evolves in time according to
    \be
    \fc(\epsilon,\vec x;x^0)=
    \int d^3 y\:G_\epsilon(\vec x,\vec y)\:
    \fc(\epsilon,\vec y;x^0_0),
    \label{ftcoh-gr}
    \ee
where we have used (\ref{fmom}), (\ref{zetatcoh}), the identity
$\br\xi_{\epsilon, \vec x}|\xi_{\epsilon, \vec p}\kt = \br\vec
x|\vec p\kt = e^{i\vec p \cdot\vec x/\hbar}/(2\pi\hbar)^{3/2}$,
and introduced the kernel
    \be
    G_\epsilon(\vec x,\vec y) := \frac{1}{(2\pi\hbar)^3}\,\int d^3
    p\:e^{i\frac{\vec p}{\hbar}\cdot(\vec x - \vec y)}\:
    e^{-i\epsilon(\frac{x^0-x^0_0}{\hbar})\sqrt{\vec p^2+m^2c^2}}.
    \label{green-ep}
    \ee
Note that because ${\cal Q}'=e\sigma_3$ commutes with $H'$,
$|\vec\zeta_\epsilon,\epsilon;x^0\kt$ has a definite charge for
all $x^0\in\R$.

\subsection{Coherent Klein-Gordon Fields}

The coherent KG fields $\psi_a^{(\epsilon,\vec\zeta_\epsilon)}$
that belong to the Hilbert space ${\cal H}_a$ are related to the
two-component coherent state vectors
$|\vec\zeta_\epsilon,\epsilon\kt$ according to (\ref{z-epsilon}).
Using this equation and (\ref{U-a-inv}), we have\footnote{In the
nonrelativistic limit, $c\rightarrow\infty$, where
$D^{-1/4}\to\cum^{-1/2}+\frac{1}{4}\cum^{-5/2}\nabla^2$,
$\psi_a^{(\epsilon,\vec\zeta_\epsilon)}(x^0_0)$ tend to the
nonrelativistic coherent state vectors $|\vec\zeta_\epsilon\kt$
provided that we set $\kappa^{-1} = 1+\epsilon a$.}
    \be
    \psi_a^{(\epsilon,\vec\zeta_\epsilon)}(x^0)=
    \sqrt{\frac{\cum}{\kappa(1+\epsilon a)}}~D^{-1/4}
    e^{-i\epsilon(x^0-x^0_0)D^{1/2}}
    \:|\vec\zeta_\epsilon\kt \in L^2(\R^3).
    \label{exp-cph}
    \ee
The value of the coherent KG field at a spacetime point has the
form
   {\small\be
   \psi_a^{(\epsilon,\vec\zeta_\epsilon)}(x)\!:=\!\br\vec
   x|\psi_a^{(\epsilon,\vec\zeta_\epsilon)}(x^0)\kt\!=\!
   \left[\frac{\cum}{8\pi^3\hbar^2\kappa (1+\epsilon
   a)} \right]^{1/2}\!\!\!\!
   \int_{\mathbb{R}^3} d^3 p\,
   \frac{e^{-i\hbar^{-1}[\epsilon(x^0-x^0_0)
   \sqrt{\vec p^2+m^2 c^2}-\vec p\cdot\vec x]}}{
   [\vec p^2+m^2 c^2]^{1/4}}\fc(\epsilon,\vec p) .
   \label{psi-coh}%
   \ee}
Figs.~\ref{probt0} -- \ref{probepsm} show the graphs of the
$(1+1)$-dimensional analogs of
$|\psi_a^{(\epsilon,\vec\zeta_\epsilon)}(x)|^2$.

Since $\psi_a^{(\epsilon,\vec\zeta_\epsilon)}$ is related by a
unitary transformation to $|\vec\zeta_\epsilon,\epsilon\kt$, they
share the properties of nonorthogonality, normalizability, and
over-completeness. We can use
$\psi_a^{(\epsilon,\vec\zeta_\epsilon)}$ to yield a coherent state
representation of arbitrary KG fields $\psi\in{\cal H}_a$:
    \[\psi=\sum_{\epsilon=\pm}\int
    \frac{d^2\zeta_\epsilon}{\pi}\,g(\vec\zeta_\epsilon,\epsilon)\,
    \psi_a^{(\epsilon,\vec\zeta_\epsilon)},\]
where the wave function associated with $\psi$ in coherent state
representation is defined by $g(\vec\zeta_\epsilon,\epsilon) :=\,
(\psi_a^{(\epsilon,\vec\zeta_\epsilon)},\psi)_a$.

\section{Coherent States of a Free Charged Scalar Field in\\
 $(1+1)$-Dimensions}

In this section we examine the coherent KG fields in
$(1+1)$-dimensions in more detail. For brevity of notation we
will use the following scaled coherent state variable
   \be
    \eta_\epsilon = \sqrt{\frac{2\hbar}{k}}\,\zeta_\epsilon =
    \alpha_\epsilon+i \beta_\epsilon,
   \label{xidefn}
   \ee
where $\alpha_\epsilon:=\Re(\eta_\epsilon)$ and
$\beta_\epsilon:=\Im(\eta_\epsilon)$ have the dimension of length.

We recall that the coherent state wave function $\fc(\epsilon,x)$
may be identified with the following normalized solution of
(\ref{eket2}), \cite{cohen, merzbacher}.
   \be
    \fc(\epsilon,x) = \left[\frac{k}{\pi \hbar}\right]^{1/4}\!\!\!
    e^{-i \frac{k}{2\hbar} \alpha_\epsilon \beta_\epsilon}\,
    e^{i\frac{k\beta_\epsilon}{\hbar} x}\,
    e^{-\frac{k}{2\hbar}(x - \alpha_\epsilon)^2}.
   \label{xepcoh}
   \ee
The Fourier transform of $\fc(\epsilon,x)$ yields the
corresponding coherent state wave function in the momentum
representation (\ref{fmom}),
    \be
    \fc(\epsilon,p)
    =\left[\frac{1}{\pi \hbar k}\right]^{1/4}\!\!\!
    e^{i\frac{k}{2\hbar} \alpha_\epsilon \beta_\epsilon}\,
    e^{-i\frac{\alpha_\epsilon}{\hbar} p}\,
    e^{-\frac{1}{2\hbar k}(p - k\beta_\epsilon)^2}.
   \label{pepcoh}
   \ee

We obtain the coherent KG fields by substituting (\ref{pepcoh}) in
the $(1+1)$-dimensional analog of (\ref{psi-coh}). This yields
    {\small\be
    \psiac(x^0,x)=\fc(\epsilon,x)\sqrt{\frac{\cum\hbar}{\kappa
    (1+\epsilon a)(2\pi\hbar k)}}\int_{\mathbb{R}} d p\,
    \frac{e^{-i\epsilon(\frac{x^0-x^0_0}{\hbar})
    \sqrt{\vec p^2+m^2 c^2}}
    e^{-\frac{1}{2\hbar k}\,[p-k\beta_\epsilon-i k
    (x-\alpha_\epsilon)]^2}}{[p^2+(m c)^2]^{1/4}}.
    \label{psi-coh-1d}%
    \ee}
One can check that the nonrelativistic limit of $\psiac(x^0_0,x)$
is the nonrelativistic coherent wave function $\fc(\epsilon,x)$.

\subsection{Observables, Uncertainty Relations, and the Classical
Limit}

In order to examine the physical properties of the coherent states
constructed above we compute the expectation values of the basic
observables in a coherent state and compare the result with the
corresponding classical quantities. We also derive the associated
minimum uncertainty relations and study their time-evolution and
their nonrelativistic limit.

It is well-known that the physical quantities such as transition
amplitudes and expectation values of observables are independent
of the choice of representation of the quantum system
\cite{ap,p57}. For example consider the observables $O':{\cal
H}'\rightarrow{\cal H}'$ and $O_a=U_a^{-1} O' U_a:{\cal
H}_a\rightarrow{\cal H}_a$ which are respectively associated with
the Foldy representation and the one-component representation.
Then,
    \be
    \br O_a \kt_\epsilon := (\psi_a^{(\epsilon,\vec\zeta_\epsilon)},
    \,O_a\,\psi_a^{(\epsilon,\vec\zeta_\epsilon)})_a =
    \br\vec\zeta_\epsilon,\epsilon|O'|\vec\zeta_\epsilon,\epsilon\kt
    =:\br \,O' \kt_\epsilon,
    \label{eq-mean}
    \ee
where we have used (\ref{z-epsilon}). In computing expectation
values we will make use of these relations and the following
formulas that follow from (\ref{xepcoh}) and (\ref{pepcoh}).
    \bea
    \br r(X')\kt_\epsilon\aeq\sqrt{\frac{k}{\pi \hbar}}\,
    \int_{\mathbb{R}} d x \,
    r(x)\,e^{-\frac{k}{\hbar}(x-\alpha_\epsilon)^2},
    \label{1exphx}\\
    \br s(P')\kt_\epsilon\aeq\frac{1}{\sqrt{\pi \hbar k}}\,
    \int_{\mathbb{R}} d p \,
    s(p)\,e^{-\frac{1}{\hbar k}(p-k\beta_\epsilon)^2},
    \label{1expgp}\\
    \br s({\rm p}) \sigma_3\kt_\epsilon\aeq\frac{\epsilon}{
    \sqrt{\pi \hbar k}}\,\int_{\mathbb{R}} d p \,
   s(p)\,e^{-\frac{1}{\hbar k}(p-k\beta_\epsilon)^2},
   \label{1expgp3}
   \eea
where $r$ and $s$ are arbitrary functions rendering the integrals
in (\ref{1exphx}) -- (\ref{1expgp3}) convergent.

To explore the `coherence' behavior of the solution given in
Eq.~(\ref{xepcoh}) we first examine the minimum uncertainty
relationship. In view of (\ref{1exphx}) and (\ref{1expgp}), we can
easily show that
   \bea
    \br X\kt_\epsilon\aeq\br X'\kt_\epsilon=
    \alpha_\epsilon\,, ~~~~~~~~~~~~~~
    \br X^2\kt_\epsilon=\br X'^2\kt_\epsilon =
    \alpha_\epsilon^2 + \frac{\hbar}{2k}\,,
    \label{x0e}\\
    \br P\kt_\epsilon\aeq\br P'\kt_\epsilon=
    k\,\beta_\epsilon\,, ~~~~~~~~~~~~
    \br P^2\kt_\epsilon=\br P'^2\kt_\epsilon =
    k^2\,\beta_\epsilon^2 + \frac{\hbar k}{2},
    \label{p0e}
    \eea
where $\alpha_\epsilon$ and $\beta_\epsilon$ are introduced in
(\ref{xidefn}). In view of (\ref{x0e}) and (\ref{p0e}), we have
the following dispersions for position and momentum operators,
respectively.
   \bea
    (\Delta X)_\epsilon^2\aeq(\Delta X')_\epsilon^2
    :=\br X'^2\kt_\epsilon - \br X'\kt_\epsilon^2 =
    \frac{\hbar}{2k}\,, \label{0xdisp}\\
    (\Delta P)_\epsilon^2\aeq(\Delta P')_\epsilon^2
    :=\br P'^2\kt_\epsilon - \br P'\kt_\epsilon^2 =
    \frac{\hbar k}{2}\,.
    \label{0disp}
   \eea
Hence, as expected, $(\Delta X)_\epsilon (\Delta P)_\epsilon =
(\Delta X')_\epsilon (\Delta P')_\epsilon =\hbar/2$, i.e., the
minimum uncertainty relation is realized by the initial coherent
state (at $x^0=x^0_0$).

Next, we examine the effect of time-evolution on the minimum
uncertainty relation. The evolution of the operators $X'$ and $P'$
in the Heisenberg picture are determined by
   \be
    \frac{d X'}{d x^0}=\frac{1}{i\hbar}\,[X', H'] =
    \frac{{\rm p}}{\sqrt{{\rm p}^2 + m^2 c^2}}\,\sigma_3\,,~~~~~~~~
    \frac{d P'}{d x^0}=\frac{1}{i\hbar}\,[P', H'] = 0\,.
    \label{xevol}
   \ee
The Heisenberg operators $X'(x^0)$ and $P'(x^0)$ are therefore
given by
  \be
  X'(x^0)= X' + \frac{ x^0\,{\rm p}}{\sqrt{{\rm p}^2 + m^2 c^2}}\,
  \sigma_3\,,~~~~~~~~~~~~~
  P'(x^0)= P'.
  \label{xt}
  \ee
Using (\ref{xevol}), (\ref{xt}) and (\ref{1expgp3}) and doing the
necessary calculations, we have
   \bea
   \br X'(x^0)\kt_\epsilon &=&
   \br X'\kt_\epsilon + \br\frac{d X'}{d x^0}\kt_\epsilon\,x^0\,,
   \label{1xtexp}\\
   \br\dot X'\kt_\epsilon &=& \br\frac{d X'}{d x^0}\kt_\epsilon =
   \frac{\epsilon}{\sqrt{\pi \hbar k}}\,\int_{\mathbb{R}} d p \,
   \frac{{\rm p}}{\sqrt{{\rm p}^2 + m^2 c^2}}\,
   e^{-\frac{1}{\hbar k}(p-k\beta_\epsilon)^2},
   \label{1xdot1}\\
   \br X'(x^0)^2\kt_\epsilon
   &=&\br X'^2\kt_\epsilon +
   2\alpha_\epsilon \br\dot X'\kt_\epsilon\,x^0 +
   \br \dot X'^2\kt_\epsilon\,x^{0\,2},
   \label{1x2t}\\
   \br\dot X'^2\kt_\epsilon &=&
   \br\left( \frac{d X'}{d x^0} \right)^2\kt_\epsilon =
   \frac{1}{\sqrt{\pi \hbar k}}\,\int_{\mathbb{R}} d p \,
   \frac{p^2}{p^2 + m^2 c^2}\,e^{-\frac{1}{\hbar k}
   (p-k\beta_\epsilon)^2}.
   \label{xdot2}
   \eea
Now, we employ (\ref{1xtexp}) and (\ref{1x2t}) to compute the
dispersion of the position operator $X'(x^0)$ in the coherent
state:
   \be
   (\Delta X'(x^0))_\epsilon^2=\frac{\hbar}{2 k} + (\Delta \dot X')_\epsilon^2\,x^{0\,2} ,
   \label{xdisp}
   \ee
where $(\Delta \dot X')_\epsilon^2:=\br\dot X'^2\kt_\epsilon -
\br\dot X'\kt_\epsilon^2$. Since $P'$ is conserved, the dispersion
$(\Delta P')_\epsilon$ does not change in time, and for all
$x^0\in\R$
   \be
   (\Delta X(x^0))_\epsilon (\Delta P(x^0))_\epsilon=
   (\Delta X'(x^0))_\epsilon (\Delta P'(x^0))_\epsilon =
   \frac{\hbar}{2} \sqrt{
   1 + \frac{2 k}{\hbar}\,(\Delta \dot X')_\epsilon^2\,x^{0\,2}} .
   \label{tunc}
   \ee

The integrals in (\ref{psi-coh-1d}), (\ref{1xdot1}) and
(\ref{xdot2}) cannot be evaluated analytically, and we could not
obtain the explicit form of $\psiac(x^0_0,x)$, $\br\dot
X'\kt_\epsilon$, $\br\dot X'^2\kt_\epsilon$ and $(\Delta \dot
X')_\epsilon^2$. We have instead calculated them numerically and
plotted their graphs as functions of the expectation value of
momentum ($k\beta_\epsilon$), time, and other relevant parameters.
To describe the behavior of these quantities, it is convenient to
introduce the dimensionless parameter $\lambda:=\sqrt{2\hbar k}/(m
c)$ and express the relevant relations in dimensionless units,
i.e., the units where the momentum, position, energy and time are
respectively measured in units of $mc$, $\lambda_c=\hbar/(m c)$
(the Compton wavelength), $m c^2$ and $\lambda_c/c=\hbar/(m c^2)$.
Note that in our notation the quantity $\dot X=\frac{d X}{c d t}$
is dimensionless, so the velocity is measured in units of $c$.

The dimensionless parameter $\lambda$ is just the ratio of the
Compton wavelength and the characteristic oscillator length
$\sigma:=\sqrt{\hbar/(2 k)}$, i.e., $\lambda:=\lambda_c/\sigma$.
In view of (\ref{0xdisp}), $\sigma$ determines the width of the
wave packet, and $\lambda^{-1}$ is a measure of the localization
of the wave packet. It is usually argued that a relativistic
particle cannot be localized more accurately than the Compton
wavelength, for otherwise pair production occurs for $E > 2 m
c^2$, \cite{greiner}.\footnote{For a critical assessment see
\cite{bracken} and references therein.} This situation restricts
the range of allowed values of $\lambda$. If we rewrite
(\ref{0xdisp}) in dimensionless unit, we find that the condition
$\Delta X\geq\lambda_c$ implies $\lambda \leqslant 1$. We will see
from our plots that the smaller the value of $\lambda$ becomes the
more classical the coherent state behaves.\footnote{Lev et al
\cite{lev-colab} use very sharply localized states in their
graphs. For example in some of their plots they take
$\lambda=50$!}

Expressing $\br\dot X'\kt_\epsilon$, $\br\dot X'^2\kt_\epsilon$,
$\br X'(x^0)\kt_\epsilon$, $(\Delta X'(x^0))_\epsilon^2$ and
$(\Delta X(x^0))_\epsilon (\Delta P(x^0))_\epsilon$ in
dimensionless units yields
   \bea
   \br\dot X'\kt_\epsilon\aeq\frac{\epsilon}{\lambda}\,
   \sqrt{\frac{2}{\pi}}\,\int_{\mathbb{R}} d p \,
   \frac{{\rm p}}{\sqrt{1 + {\rm p}^2}}\,e^{-\frac{2}{\lambda^2}
   ( p - k\beta_\epsilon)^2},
   \label{1xdot11}\\
   \br\dot X'^2\kt_\epsilon\aeq\frac{1}{\lambda}\,
   \sqrt{\frac{2}{\pi}}\,\int_{\mathbb{R}} d p \,
   \frac{{\rm p}^2}{1 + {\rm p}^2}\,e^{-\frac{2}{\lambda^2}
   ( p - k\beta_\epsilon)^2},
   \label{xdot22}\\
   \br X'(\tau)\kt_\epsilon\aeq\br X'\kt_\epsilon +
   \br\dot X'\kt_\epsilon\,\tau\,,
   \label{1xtexp11}\\
   (\Delta X'(\tau))_\epsilon\aeq(\Delta X(\tau))_\epsilon =
   \frac{1}{\lambda}
   \sqrt{1 + \lambda^2 (\Delta \dot X')^2_\epsilon\,\tau^2}\,,
   \label{xdisp2}\\
   (\Delta X(\tau))_\epsilon (\Delta P(\tau))_\epsilon\aeq(
   \Delta X'(\tau))_\epsilon (\Delta P'(\tau))_\epsilon =
   \frac{1}{2} \sqrt{1 + \lambda^2 (\Delta \dot X')^2_\epsilon\,\tau^2}\,,
   \label{tunc2}
   \eea
where $\tau:=x^0/\lambda_c=c t/\lambda_c$ is a dimensionless time
parameter. The nonrelativistic quantum mechanical analogs of
(\ref{1xdot11}) -- (\ref{tunc2}) have the form
    \bea
   \br\dot{\rm x}\kt_{\rm nr}\aeq k\beta = \br{\rm p}\kt,~~~~~~
   \br\dot{\rm x}^2\kt_{\rm nr} =\frac{\lambda^2}{4} + \br{\rm p}\kt^2,
   ~~~~~~
   \br{\rm x}(\tau)\kt_{\rm nr}=\br{\rm x}\kt + \br{\rm p}\kt\,
   \tau\,, \label{nrxtexp}\\
   \Delta{\rm x}(\tau)\aeq\frac{1}{\lambda}\sqrt{1 +
   \frac{1}{4} \lambda^4 \tau^2}\,,~~~~~~
   \Delta{\rm x}(\tau)\Delta{\rm p}(\tau)=\frac{1}{2}
   \sqrt{1 + \frac{1}{4} \lambda^4 \tau^2}\,.
   \label{nrtunc}
   \eea
Now, we are in a position to plot these quantities and perform a
relativistic-to-nonrelativistic and quantum-to-classical
comparisons.

Fig.~\ref{xdotfig} shows the dependence of the velocity
expectation value $\br\dot{\rm x}\kt$ and dispersion in velocity
$\Delta \dot{\rm x}$ on momentum expectation value $\br{\rm p}\kt$
for different values of $\lambda$. It also includes the graph of
the corresponding classical curve, i.e., $\br\dot{\rm
x}\kt=\frac{\br{\rm p}\kt}{\sqrt{1+\br{\rm p}\kt^2}}$. As one
reduces the value of $\lambda$, the quantum curve tends to the
classical curve. Moreover, for higher velocities the dispersion in
velocity is small, and as we shall argue below, the spreading of
the wave packet is slower. Similarly, the dispersion in position
(the width of the coherent wave packet), the product of the
position and momentum dispersions, and consequently the minimum
uncertainty relation are velocity-dependent. This is a feature of
the relativistic coherent states that does not survive the
nonrelativistic limit; in nonrelativistic QM these quantities are
velocity-independent.

Fig.~\ref{dxfig} shows the dispersion in position as a function of
time for various values of the momentum expectation value and the
parameter $\lambda$. It also shows the corresponding
nonrelativistic quantum mechanical dispersion. The relativistic
dispersion turns out to be smaller than the nonrelativistic
dispersion. Also as one increases the momentum the uncertainty in
position decreases. In view of the fact that the momentum operator
does not depend on time, this shows that the product of the
position and momentum dispersions is smaller for the faster moving
packets. See Figs.~\ref{dxdpfig} which also show that the
spreading of a faster moving coherent wave packet has a smaller
rate than that of the slower moving packets. We will arrive at the
same conclusion when we consider the graphs of the evolution of
the probability density below.

Next, we explore the behavior of the energy expectation value.
Recall that in the Foldy representation the Hamiltonian operator
is given by $H' = \sqrt{{\rm p}^2 + m^2 c^2}\,\sigma_3$. Hence, we
define the energy operator according to
     \be
     E':=c\, {\cal C}' H' = \sqrt{c^2{\rm p}^2 + m^2
     c^4}\,\sigma_0\,.
     \label{energy-op}
     \ee
In view of (\ref{1expgp3}), its expectation value is given in
dimensionless units by
   \be
   \br E\kt_\epsilon := \br E'\kt_\epsilon = \br\sqrt{c^2 {\rm p}^2 + m^2 c^4}\kt_\epsilon =
   \frac{1}{\lambda}\,\sqrt{\frac{2}{\pi}}\,\int_{\mathbb{R}} d p \,
   \sqrt{1 + {\rm p}^2}\,e^{-\frac{2}{\lambda^2}( p - k\beta_\epsilon )^2}.
   \label{1energy2}
   \ee
We can also compare it with its nonrelativistic counterpart,
namely
   \be
   \br E\kt_{\rm nr} = 1 + \br\frac{{\rm p}^2}{2}\kt =
   1 + \frac{1}{\lambda}\,
   \sqrt{\frac{2}{\pi}}\,\int_{\mathbb{R}} d p \,{\rm p}^2\,
   e^{-\frac{2}{\lambda^2}( p - k\beta_\epsilon )^2}
   = 1 + \frac{\lambda^2}{8} + \frac{\br{\rm p}\kt^2}{2} .
   \label{nonrelenrg}
   \ee

In Fig.~\ref{energy} we plot the expectation value of energy and
its dispersion $\Delta E$ in terms of the momentum expectation
value for different $\lambda$ and the corresponding classical
(nonquantum) and nonrelativistic (quantum mechanical) curves. From
these graphs we can see that for small values of $\lambda$ the
coherent states display completely classical behavior.
Table~\ref{tab1n} shows some typical relativistic and
nonrelativistic energy expectation values for small momentum and
various values of $\lambda$. The relativistic expectation value of
energy is smaller than its nonrelativistic counterpart and closer
to the classical relativistic energy. For small values of
$\lambda$ the relativistic and nonrelativistic energy expectation
values coincide and tend to the classical result.
  \begin{table}[h]
    \begin{center}\begin{tabular}{||p{1.94cm}||p{1.78cm}p{1.66cm}p{1.91cm}||p{1.94cm}p{2cm}p{2.34cm}||}
    \hline\hline
    \multicolumn{1}{||c||}{$\br{\rm p}\kt$} &  \multicolumn{1}{c}{} &  \multicolumn{1}{c}{$0.1$} &  \multicolumn{1}{c||}{} &
    \multicolumn{1}{c}{} &  \multicolumn{1}{c}{$0.001$} &  \multicolumn{1}{c||}{} \\
    \hline
    \multicolumn{1}{||c||}{$\lambda$} &  \multicolumn{1}{c|}{$0.25$} &  \multicolumn{1}{c|}{$0.5$} &
    \multicolumn{1}{c||}{$2.0$} &  \multicolumn{1}{c|}{$0.25$} &  \multicolumn{1}{c|}{$0.5$} &  \multicolumn{1}{c||}{$2.0$} \\
    \hline\hline
    \multicolumn{1}{||c||}{$\br E\kt/E_{\rm cl.}$} &  \multicolumn{1}{c|}{$1.00758$} &  \multicolumn{1}{c|}{$1.02944$} &
    \multicolumn{1}{c||}{$1.35062$} & \multicolumn{1}{c|}{$1.00772$} &  \multicolumn{1}{c|}{$1.02997$} &  \multicolumn{1}{c||}{$1.35453$} \\
    \hline
    \multicolumn{1}{||c||}{$\br E\kt_{\rm nr}/E_{\rm cl.}$} &  \multicolumn{1}{c|}{$1.00777$} &  \multicolumn{1}{c|}{$1.03109$} &
    \multicolumn{1}{c||}{$1.49751$} & \multicolumn{1}{c|}{$1.00781$} &  \multicolumn{1}{c|}{$1.03125$} &  \multicolumn{1}{c||}{$1.50000$} \\
    \hline\hline
    \end{tabular}\end{center}
    \centerline{
    \parbox{14cm}{
    \caption{Comparison of the relativistic and nonrelativistic
    energy expectation values for small momenta. The
    relativistic expectation value is smaller than the
    nonrelativistic expectation value and closer to the classical
    relativistic energy. For small values of $\lambda$ the
    relativistic and nonrelativistic energy expectation values
    coincide and tend to the classical result.}
    \label{tab1n}}}
    \end{table}

\subsection{Probability Density and Its Time-Evolution}

The probability density $\rho(x^0,x)$ for a coherent state with
definite charge parity $\epsilon$, in dimensionless units, has the
form
   \be
   \rho(\epsilon;x^0,x) := |\fc(\epsilon; x^0,x)|^2 =
   \frac{1}{\lambda\pi\sqrt{2\pi}}\left[S^2 + T^2\right],
   \label{prob-rel}
   \ee
where we have made use of Eqs.~(I-63), (\ref{ftcoh-gr}),
(\ref{green-ep}), (\ref{xepcoh}) and introduced
   \bea
   S\adef\int_{\mathbb{R}} d p\,
   e^{-\frac{1}{\lambda^2}( p-k\beta_\epsilon )^2}
   \,\cos{\left[ p (x-\alpha_\epsilon) -
   \epsilon\tau\sqrt{1+p^2}\right]},\\
   T\adef\int_{\mathbb{R}} d p\,
   e^{-\frac{1}{\lambda^2}( p-k\beta_\epsilon )^2}
   \,\sin{\left[ p (x-\alpha_\epsilon) -
   \epsilon\tau\sqrt{1+p^2}\right]}.
   \eea
Evaluating these integrals or using Eq.~(\ref{xepcoh}) yields
   \be
   \rho(\epsilon;x^0_0,x) = |\fc(\epsilon; x^0_0,x)|^2 =
   \frac{\lambda}{\sqrt{2\pi}}\,
   e^{-\frac{\lambda^2}{2} (x - \alpha_\epsilon)^2} .
   \ee
Therefore, the probability density at time $x^0_0$ is just the
nonrelativistic Schr\"odinger probability density and does not
depend on the particle's momentum. But at later times, the
probability density depends on the particle's momentum $\br{\rm
p}\kt_\epsilon=k\beta_\epsilon$. Also, from these equations we see
that for $\tau\neq 0$ the probability density depends on
$\epsilon$, hence its evolution differs for positive and negative
charges. The difference is that the packets with negative charge
parity move backward in time. This means that the physical
momentum is to be identified with $\epsilon\, {\rm p}$,
\cite{Gitman}. Figs.~\ref{probt0} -- \ref{probepsm} show plots of
$\rho(\epsilon;x^0,x)$.

Eq.~(\ref{psi-coh-1d}) gives an expression for the coherent KG
field, $\psiac$, in $(1+1)$-dimensions. In order to compare this
expression with that of its nonrelativistic analog, namely
$\fc(\epsilon; x^0,x)$, we plot them in terms of the momentum
expectation value for various values of the parameter $\lambda$.
First, we rewrite (\ref{psi-coh-1d}) in dimensionless units and
compute
   \be
   |\psiac(x^0,x)|^2 = \frac{1}{\lambda \pi \kappa
   (1+\epsilon a)\sqrt{2\pi}}
   \left[U^2 + V^2\right],
   \ee
where
   \bea
   U\adef\int_{\mathbb{R}} d p\,\frac{e^{-\frac{1}{\lambda^2}
   ( p-k\beta_\epsilon )^2}}
    {[1 + p^2]^{\frac{1}{4}}}\,\cos{\left[ p (x-\alpha_\epsilon) -
    \epsilon\tau\sqrt{1+p^2}\right]},\nn\\
   V\adef\int_{\mathbb{R}} d p\,\frac{e^{-\frac{1}{\lambda^2}
   ( p+k\beta_\epsilon )^2}}
    {[1 + p^2]^{\frac{1}{4}}}\,\sin{\left[ p (x-\alpha_\epsilon) -
    \epsilon\tau\sqrt{1+p^2}\right]}.\nn
   \eea
In the nonrelativistic limit, $U$ and $V$ respectively tend to $S$
and $T$. This means that the nonrelativistic limit of
$|\psiac(x^0,x)|^2$ coincides with the nonrelativistic probability
density $|\fc(\epsilon; x^0,x)|^2$ for all $x^0\in\R$.

Fig.~\ref{probt0} gives the plots of $|\fc(\epsilon;x^0_0,x)|^2$
and $|\psiac(x^0_0,x)|^2$ for various values of the momentum
expectation value, $\br p\kt_\epsilon=k\beta_\epsilon$, and
$\lambda$. As seen from this figure, $|\psiac(x^0_0,x)|^2$ depends
on the particle's initial momentum expectation value. For small
values of the latter $|\psiac(x^0_0,x)|^2$ tends to
$|\fc(\epsilon;x^0_0,x)|^2$.

Fig.~\ref{probtp1} shows that both the probability density
$\rho(\epsilon; x^0,x)$ and modulus-square $|\psiac(x^0,x)|^2$ of
a coherent KG wave packet spread with time while their maximum
value decreases. The nature of this spreading depends on the
momentum expectation value and the initial width of the wave
packet. The more localized the initial coherent packet is (the
closer $\lambda$ gets to one) the faster it spreads. This is in
complete accordance with Figs.~\ref{dxfig} and \ref{dxdpfig}.
Also, the fast moving packets behave more like a classical
particle; they travel a larger distance without noticeable
spreading.

The spreading of the relativistic coherent wave packet is similar
to the one encountered in nonrelativistic QM \cite{VQM}. However,
unlike for a nonrelativistic coherent wave packet, the maximum of
the probability density $\rho(\epsilon; x^0,x)$ and the
modulus-square $|\psiac(x^0,x)|^2$ of a relativistic coherent wave
packet do not move with its velocity expectation value or the
velocity of the corresponding classical particle. They move with a
higher velocity which is nevertheless smaller than the velocity of
light. This is easily seen from the graphs given in
Fig.~\ref{probtp1}.\footnote{For example in these graphs, for
$\lambda=1.0$ and $\br{\rm p}\kt=1.0$, the successive maxima of
$\rho(\epsilon; x^0,x)$ are $x=0.0, 7.4, 15.5, 23.7, 31.8$ and
those of $|\psiac(\epsilon; x^0_0,x)|^2$ are $x=0.0, 7.1, 15.0,
23.0, 30.9$. For $\lambda=0.5$ and $ \br{\rm p}\kt=1.0$, the
successive maxima of $\rho(\epsilon; x^0,x)$ are $x=0.0, 10.4,
21.4, 32.6, 43.6$ and those of $|\psiac(\epsilon; x^0_0,x)|^2$ are
$x=0.0, 10.2, 21.1, 32.2, 43.2$. For $\lambda=1.0$ and $\br{\rm
p}\kt=2.0$ the successive maxima of $\rho(\epsilon; x^0,x)$ are
$x=0.0, 27.0, 54.5, 82.0, 109.6$ and those of $|\psiac(\epsilon;
x^0_0,x)|^2$ are $x=0.0, 26.8, 54.1, 81.5, 108.9$. The velocity of
the corresponding classical particle is $\frac{\br{\rm
p}\kt}{\sqrt{1+\br{\rm p}\kt^2}}=0.7071$ and $0.8944$ respectively
for $\br{\rm p}\kt=1.0$ and $2.0$, while the quantum expectation
values of velocity $\br\dot{\rm x}\kt$ are respectively $0.6421,
0.6903$ and $0.8786$ for the three cases: (1) $\lambda=1.0,
\br{\rm p}\kt=1.0$; (2) $\lambda=0.5, \br{\rm p}\kt=1.0$; (3)
$\lambda=1.0, \br{\rm p}\kt=2.0$.\label{foot7}}

Fig.~\ref{probepsm} shows the time-evolution of the probability
density and $|\psiac(x^0_0,x)|^2$ for $\epsilon=-1$ and a positive
initial momentum expectation value. As demonstrated by this graph,
the probability density and the KG wave function evolve in $-x^0$
direction (in contrast to Fig.~\ref{probtp1}).

\section{Coherent States of a Free Neutral Field}

Neutral scalar particles are described by real KG fields which we
briefly studied in \cite{p59}. As we showed there, a
characteristic feature of a real KG field is that its position
wave function $f(\epsilon,\vec x)$ satisfies $f(\epsilon,\vec
x)=f(-\epsilon,\vec x)^*$. This condition together with
(\ref{xepcoh}) restrict the parameters $\beta_\epsilon$ and
$\alpha_\epsilon$ of the coherent states according to
    \be
    \beta_{-\epsilon} = - \beta_\epsilon,~~~~~~~~~~
    \alpha_{-\epsilon} = \alpha_\epsilon .
    \label{param-con}
    \ee
In the one-component representation, the coherent states for
neutral field is defined by
    \be
    \psi_a^{(\vec\zeta)}:= \frac{1}{\sqrt 2}\left[
    \psi_a^{(+,\vec\zeta)} + \psi_a^{(-,\vec\zeta^*)} \right],
    \ee
where the coherent wave functions $\fc(+,\vec x)$ and $\fc(-,\vec
x)$ are given by (\ref{xepcoh}) and (\ref{param-con}).

As seen from Fig.~\ref{probepsm} the coherent wave packet with
negative charge parity moves backward in time. Therefore, the
physical momentum for states with charge parity $\epsilon$ is
$\epsilon \vec{\rm p}$, \cite{Gitman}. Hence, the physical
momentum operator for a general KG field may be identified with
${\cal C}\vec{\rm p}$. Using this expression in computing the
expectation values, probability densities, etc., we have found
that all the results and graphs of the preceding section are valid
for a free neutral particle.

\section{Coupling to a Uniform Magnetic Field}

Consider a scalar charged particle with the electric charge $e$ in
a constant homogeneous magnetic field directed along the
(positive) $x^3$-axis, $\vec B=(0,0,B)$ with $B>0$. In the
symmetric gauge the electromagnetic vector potential has the form
   \be
   \vec A(\vec x) = - \frac{1}{2} ( \vec x \times \vec B )
   ~~\Rightarrow~~~~
   A_0 = A_3 = 0, \hspace{0.5cm}A_1 = -
   \frac{1}{2} B x^2, \hspace{0.5cm}A_2 = \frac{1}{2} B x^1
   \label{vecpot}
   \ee

In order to fix the notation and compare the quantum mechanical
and classical results, we first review the classical treatment of
the problem.

\subsection{Classical Treatment}

The classical motion may be obtained using the classical
Hamiltonian \cite{jackson}
   \be
   H_{\rm cl.} = E:=\sqrt{ c^2 \vec\Pi^2 + m^2 c^4 },
   \label{mf-ch}
   \ee
where $E$ and $\vec\Pi$ stand for the particle's energy and
kinetic momentum, respectively. The latter is given, in the gauge
(\ref{vecpot}), by
   \be
   \vec\Pi = \vec p - e \vec A = \gamma m \vec v,\hspace{0.75cm}
   \Pi_1 = p_1 + \frac{e B}{2} x^2, \hspace{0.75cm}
   \Pi_2 = p_2 - \frac{e B}{2} x^1, \hspace{0.75cm}
   \Pi_3 = p_3,
   \label{pken1}
   \ee
where $\vec p$ is the canonical momentum conjugate to the position
$\vec x$ of the particle, $\vec v=(\dot x^1, \dot x^2, \dot x^3)$
is the velocity, and $\gamma$ is the Lorentz factor. The
Hamiltonian is a constant of motion, so the magnitude of the
velocity does not change and $\gamma$ is a constant.

The Hamilton's equations of motion read
   \bea
   \dot{\vec x}\adef\frac{d\vec x}{d x^0} =
   \frac{\vec\Pi}{\sqrt{\vec\Pi^2 + (m c)^2}}\,,
   \label{clas-vel} \\
   \dot p_1\adef\frac{d p_1}{d x^0} =
   \frac{(\frac{e B}{2})\Pi_2}{\sqrt{\vec\Pi^2 + (m c)^2}}\,,~~~~~~
   \dot p_2 := \frac{d p_2}{d x^0} =
   \frac{- (\frac{e B}{2})\Pi_1}{\sqrt{\vec\Pi^2 + (m c)^2}}\,, ~~~~~~
   \dot p_3 := \frac{d p_3}{d x^0} = 0.
   \label{pconj}
   \eea
They are equivalent to \cite{jackson}
    \be
   \frac{d \vec v}{d x^0} = \vec v \times \vec\omega_B ,
   \label{vdot-mg}
   \ee
where $\vec\omega_B$ is the gyration or precession
frequency\footnote{Note that in our notation $\vec\omega_B$ has
the unit (length)$^{-1}$. The physical frequency is $c
|\vec\omega_B|$.} and has the form
   \be
   \vec\omega_B := \frac{e \vec B}{\gamma m c} =
   \frac{e c \vec B}{E} .
   \label{gyr-freq}
   \ee

The motion described by (\ref{vdot-mg}) is a circular motion
perpendicular to $\vec B$ and a uniform translational motion
parallel to $\vec B$. The solution of (\ref{vdot-mg}) has the form
\cite{jackson}
   \be
   \vec v(x^0) = \dot x^3 \hat{e}_3 + R \omega_B (\hat{e}_1 -
   i \hat{e}_2) e^{-i \omega_B x^0} ,
   \label{vsolut}
   \ee
where $R$ is the gyration radius\footnote{Note that in our
notation the product $R \omega_B$ is a dimensionless velocity.},
$\hat{e}_i$ is the unit vector along the $x^i$-axis, and the
physical velocity of the particle is given by the real part of
this equation. For a positive charge $e$, it represents a
counterclockwise rotation when viewed in the direction of $\vec
B$, \cite{jackson}. Integrating (\ref{vsolut}) yields the position
of the particle:
   \be
   \vec X(x^0) = \vec X_{\rm g.c.} + \dot x^3 x^0 \hat{e}_3 +
   i R (\hat{e}_1 - i \hat{e}_2) e^{-i \omega_B x^0} .
   \label{xsolut}
   \ee
The classical path is a helix with center $\vec X_{\rm g.c.}$,
radius $R$, and pitch angle $\vartheta\!=\!\tan^{-1}(\rp_3/e B
R)$.

In view of (\ref{clas-vel}) and (\ref{pconj}), the coordinates of
the gyration center,
   \be
   x^1_{\rm g.c.} := \frac{x^1}{2} + \frac{p_2}{e B} , ~~~~~~~~~~~
   x^2_{\rm g.c.} := \frac{x^2}{2} - \frac{p_1}{e B} ,
   \label{gyr-cent}
   \ee
are constants of motion \cite{johnson}. We can use them to express
the gyration radius $R$ and the component $L_3$ of the angular
momentum along the $x^3$-direction. This yields
   \bea
   R \aeq \sqrt{(x^1-x^1_{\rm g.c.})^2 + (x^2-x^2_{\rm g.c.})^2} =
   \frac{\Pi_\perp}{e B}\,
   \label{gyr-rad}\\
   L_3 \aeq x^1 p_2 - x^2 p_1 = \frac{e B}{2}
   \left [ (x^1_{\rm g.c.})^2 + (x^2_{\rm g.c.})^2 - R^2 \right ]
   = \frac{e B}{2} \left( R_{\rm g.c.}^2 - R^2 \right) ,
   \label{angm}
   \eea
where $\Pi_\perp := \sqrt{\Pi_1^2 + \Pi_2^2}$. These relations are
consistent with the fact that $\Pi_3$ and the total energy are
also conserved quantities \cite{jackson,johnson}.

\subsection{Quantum Mechanical Treatment}

In relativistic quantum mechanics, the charged scalar particle
interacting with a constant homogeneous magnetic field may be
described by the KG equation (\ref{kg}) where $D$ is given by
   \be
   D = \hbar^{-2} [ \vec\Pi^2 + (m c)^2 ] =
   \hbar^{-2} [ (\vec{\rm p} - e \vec A)^2 + (m c)^2 ],
   \label{mg-dop}
   \ee
and $\vec A$ is the vector potential (\ref{vecpot}).

In the Foldy representation, the Hamiltonian has the form
   \be
   H' = \sqrt{ \vec\Pi^2 + m^2 c^2 }
   \;\sigma_3 = \sqrt{ (\vec{\rm p} - e\vec A)^2 + m^2 c^2 }
   \;\sigma_3 .
   \label{mg-qh}
   \ee
In the following we shall drop the symbol ``$\otimes\,\sigma_0$''
in (\ref{xp-prime}) for simplicity, so that the position and
momentum operators in the Foldy representation are identified with
ordinary position and momentum operators acting in $L^2(\R^3)$,
i.e., $\vec X'\equiv \vec{\rm x}$ and $\vec P'\equiv \vec{\rm p}$.
Using this convention, we may express the Heisenberg equations of
motion in Foldy representation as
   \be
   \dot{\romx}=\frac{d{\mbox{$\vec{\rm x}$}}}{d x^0} =
   \frac{1}{i\hbar} \left[{\mbox{$\vec{\rm
   x}$}},H'\right]\,,~~~~~~~~~
    {\mbox{$\dot{\vec{\rm p}}$}}=\frac{d{\mbox{$\vec{\rm p}$}}}{d x^0} =
   \frac{1}{i\hbar} \left[{\mbox{$\vec{\rm p}$}},H'\right]\,.
   \label{h-p}
   \ee
In contrast to the classical case, the commutators appearing in
these equation do not admit a closed-form expression in terms of
$\vec{\rm x}$ and $\vec{\rm p}$. This makes an explicit solution
of (\ref{h-p}) intractable.\footnote{Some authors \cite{briegel}
report a symmetrization problem in equations (\ref{h-p}) which we
think is not relevant.} We have performed a first-order
perturbative\footnote{$B$ is the perturbation parameter.}
investigation of (\ref{h-p}) and shown that similarly to the
classical case the trajectory is a helix with constant gyration
center and radius. We will not include the details of this
investigation here. Rather we will present a nonperturbative
numerical treatment of the problem that turns out to be more
efficient for larger values of $\Lambda$ (respectively $B$).

In the Foldy-representation, the three-dimensional coherent state
vectors are the tensor product of one-dimensional coherent state
vectors $|\zeta^i_\epsilon,\epsilon\kt$ (with $i=1,2,3$) along the
$x^i$-directions, i.e.,
   \be
   |\vec\zeta_\epsilon,\epsilon\kt :=
   |\zeta^1_\epsilon,\epsilon\kt\otimes|\zeta^2_\epsilon,
   \epsilon\kt\otimes|\zeta^3_\epsilon,\epsilon\kt.
   \label{3d-coh}
   \ee
Because of the symmetry in $x^1$-$x^2$ plane, we take the width of
the wave packet in $x^1$ and $x^2$ directions to be equal. The
coherent state wave functions in the $x$- and $p$-representations
are respectively given by
   \bea
   \fc(\epsilon,\vec x)\aeq\fc(\epsilon, x^1)
   \fc(\epsilon, x^2) \fc(\epsilon, x^3) \nn\\
   \aeq\left[\frac{k_\perp^2 k_3}{\pi^3 \hbar^3}\right]^{1/4}
   e^{-i\frac{\vec q_\epsilon}{2\hbar}\cdot\vec\alpha_\epsilon}\,
   e^{i\frac{\vec q_\epsilon}{\hbar}\cdot\vec x}
   \,e^{-\frac{k_\perp}{2\hbar}(x^1-\alpha_{1\epsilon})^2}
   e^{-\frac{k_\perp}{2\hbar}(x^2-\alpha_{2\epsilon})^2}
   e^{-\frac{k_3}{2\hbar}(x^3-\alpha_{3\epsilon})^2},
   \label{3x-coh}\\
   \fc(\epsilon,\vec p)\aeq\fc(\epsilon, p_1)
   \fc(\epsilon, p_2) \fc(\epsilon, p_3) \nn\\
   \aeq\left[\frac{1}{k_\perp^2 k_3 \pi^3\hbar^3}\right]^{1/4}
   e^{i\frac{\vec q_\epsilon}{2\hbar}\cdot\vec\alpha_\epsilon}\,
   e^{-i\frac{\vec\alpha_\epsilon}{\hbar}\cdot\vec p}
   \,e^{-\frac{1}{2\hbar k_\perp}(p_1-q_{1\epsilon})^2}
   e^{-\frac{1}{2\hbar k_\perp}(p_2-q_{2\epsilon})^2}
   e^{-\frac{1}{2\hbar k_3}(p_3-q_{3\epsilon})^2},
   \label{3p-coh}
   \eea
where $\vec q_\epsilon:=\sum_{i=1}^3
k_i\beta_{i\epsilon}\hat{e}_i=\sum_{i=1}^3\br{\rm p}_i\kt_\epsilon
\hat{e}_i,~ k_\perp\!:=\!k_1\!=\!k_2$, and we have made use of
(\ref{xepcoh}). The wave packet's widths in $x^1$-$x^2$ and $x^3$
directions are respectively given by $\sigma_\perp=\sqrt{\hbar/(2
k_\perp)}$ and $\sigma_3=\sqrt{\hbar/(2 k_3)}$. Clearly
(\ref{3d-coh}) represents the minimum-uncertainty states.

Next, we consider a positive charged particle in such a coherent
state and drop $\epsilon=+1$ for brevity. In order to compute the
expectation value $\br\vec\zeta|\vec{\rm x}(x^0)|\vec\zeta\kt$
(resp.\ $\br\vec\zeta|\vec{\rm p}(x^0)|\vec\zeta\kt$) of the
position (resp.\ momentum) operator at any given time $x^0$, we
will use the following implicit solution of (\ref{h-p}).
 \be
 {\rm x}^i(x^0) = e^{i \frac{x^0}{\hbar} H'}
 {\rm x}^i e^{-i \frac{x^0}{\hbar} H'},~~~~~~~~~~
 {\rm p}_i(x^0) = e^{i \frac{x^0}{\hbar} H'}
 {\rm p}_i e^{-i \frac{x^0}{\hbar} H'}.
 \ee
We will also need to obtain the action of $H'$ on the coherent
state vector $|\vec\zeta\kt$. Because of the square root appearing
in (\ref{mg-qh}) a direct calculation of $H'|\vec\zeta\kt$
encounters severe problems. We will avoid them by expressing the
relevant quantities in an eigenbasis of
  \be
  h':=H'^2=(mc)^2+({\rm p}_1+\frac{eB}{2} {\rm x}^2)^2+
  ({\rm p}_2-\frac{eB}{2}
  {\rm x}^1)^2+{\rm p}_3^2\,.
  \ee
For example, for the expectation value of an operator $O(x^0)$ in
the coherent state vector $|\vec\zeta\kt$, we have
   {\small \bea
   \!\!\!\!\br\vec\zeta|O(x^0)|\vec\zeta\kt\aeq\br\vec\zeta|
   e^{i\frac{x^0}{\hbar}\sqrt{h'}}
   O e^{-i \frac{x^0}{\hbar}\sqrt{h'}}|\vec\zeta\kt\nn\\
   \aeq\sum_{n,\ell,n',\ell'} \int\! d k_3\!\!\int\!d k'_3\,
   \br\vec\zeta|e^{i \frac{x^0}{\hbar}\sqrt{h'}}
   |n,\ell,k_3\kt\br n,\ell,k_3|O|n',\ell',k'_3\kt\br n',\ell',k'_3|
   e^{-i \frac{x^0}{\hbar}\sqrt{h'}}|\vec\zeta\kt,~~~~~~
   \label{matrix-elm}
   \eea}%
where $|n,\ell,k_3\kt$ are the eigenvectors of the operators $h'$
and $L_3:={\rm x}^1 {\rm p}_2 - {\rm x}^2 {\rm p}_1$. The
corresponding normalized eigenfunctions may be obtained by solving
the eigenvalue equations:
    \[h' \psi_{(n,\ell,k_3)}(\vec x) =
    E_{(n,\ell,k_3)} \psi_{(n,\ell,k_3)}(\vec x),~~~~~~~
    L_3\psi_{(n,\ell,k_3)}(\vec x) = \ell \hbar\,
    \psi_{(n,\ell,k_3)}(\vec x).\]
They are given by
  \be
  \psi_{(n,\ell,k_3)}(\vec x) := \br\vec x|n,\ell,k_3\kt =
  \frac{(\frac{eB}{2\hbar})^{\frac{|\ell|+1}{2}}}{\sqrt{2\pi^2\hbar}}
  \left(\frac{n!}{(n+|\ell|)!}\right)^{1/2}\,e^{i \ell \varphi}
  \,e^{i \frac{k_3}{\hbar} x^3}\,\rho^{|\ell|}
  \,e^{-\frac{eB}{4\hbar} \rho^2}\,L_n^{\!|\ell|}(\frac{eB}{2\hbar}
  \rho^2) ,
  \label{eigenfunc}
  \ee
where $L_n^{\!|\ell|}(x)$ is the associated Laguerre polynomial,
$\rho$ and $\varphi$ are polar coordinates in $x^1$-$x^2$ plane,
$n=0,1,2,3,\cdots$, $\ell=0,\pm 1,\pm 2,\cdots$, and $k_3\in\R$.
The eigenvalues have the form
   \be
   E_{(n,\ell,k_3)} = (mc)^2 + k_3^2 + \hbar e B
   (2n + 1 - \ell + |\ell|).
   \label{eigenval}
   \ee

The six parameters $\alpha_{1 \epsilon}, \alpha_{2
\epsilon},\alpha_{3 \epsilon}, \beta_{1 \epsilon}, \beta_{2
\epsilon}$ and $\beta_{3 \epsilon}$, that specify the initial
expectation value of the position and momentum operators,
determine the form of the coherent state (\ref{3x-coh})
completely. Similarly to a corresponding classical particle, the
behavior of the coherent state (\ref{3x-coh}) does not depend on
its initial position and the direction of its initial momentum in
the $x^1$-$x^2$ plane. Therefore, without loss of generality, we
may consider an initial coherent wave packet that is centered at
the origin and has a momentum that lies in the $x^1$-$x^3$ plane,
i.e.,
    \be
    \alpha_{i \epsilon}=\beta_2=0,
    ~~~~~\mbox{so that}~~~~~
    \br{\rm x}^i\kt=\br{\rm p}_2\kt=0,
    \label{simpch}
    \ee
for all $i=1,2,3$ and $\epsilon=\pm$ .

Substituting (\ref{simpch}) in (\ref{3x-coh}), using
(\ref{matrix-elm}) -- (\ref{eigenval}), and performing a rather
lengthy calculation, we find in the previously introduced
dimensionless units
   {\small\bea
   \br{\rm x}^1(\tau)\kt\!\aeq\!{\cal F}
   \left\{\,\sum_{n=0}^{\infty}\sum_{\ell=0}^{\infty}
   \frac{n!}{(n+\ell)!}\,\,s^{2n+\ell} u^\ell
   L_n^{\!\ell}(u) L_n^{\!\ell+1}(u)\!
   \int_{-\infty}^{+\infty}\!\!\!\!\!\!
   d k_3\,e^{-\frac{2}{\lambda_3^2}(k_3-\br{\rm p}_3\kt)^2}
   \!\!\!\sin{[\Theta_{n\ell}(k_3)]} \right.
   \nn\\
   &&\vspace{2cm}\left.-\sum_{n=0}^{\infty}
   \sum_{\ell=0}^{\infty}\frac{(n+1)!}{(n+\ell+1)!}\,\,
   s^{2n+\ell+1} u^\ell L_{n+1}^{\!\ell}(u) L_n^{\!\ell+1}(u)
   \!\int_{-\infty}^{+\infty}\!\!\!\!\!\!d k_3\,
   e^{-\frac{2}{\lambda_3^2}(k_3-\br{\rm p}_3\kt)^2}
   \!\!\!\sin{[\Theta_{n 0}(k_3)]}\right\},
   ~~~~~~~~~\label{f-mat-x1}\\ &&\nn\\
   \br{\rm x}^2(\tau)\kt\!\aeq\!-\,\frac{\br{\rm p}_1\kt}{\Lambda} + {\cal F}
   \left\{\,\sum_{n=0}^{\infty}\sum_{\ell=0}^{\infty}
   \frac{n!}{(n+\ell)!}\,\,s^{2n+\ell} u^\ell
   L_n^{\!\ell}(u) L_n^{\!\ell+1}(u)
   \!\int_{-\infty}^{+\infty}\!\!\!\!\!\!d k_3\,
   e^{-\frac{2}{\lambda_3^2}(k_3-\br{\rm p}_3\kt)^2}\!\!\!
   \cos{[\Theta_{n\ell}(k_3)]} \right.\nn\\
   &&\vspace{2cm}\left.-\sum_{n=0}^{\infty}
   \sum_{\ell=0}^{\infty}\frac{(n+1)!}{(n+\ell+1)!}
   \,\,s^{2n+\ell+1} u^\ell L_{n+1}^{\!\ell}(u) L_n^{\!\ell+1}(u)
   \int_{-\infty}^{+\infty}
   \!\!\!\!\!\!d k_3\,e^{-\frac{2}{\lambda_3^2}(k_3-\br{\rm p}_3\kt)^2}
   \!\!\!\cos{[\Theta_{n0}(k_3)]}\right\},
   \label{f-mat-x2}\\ &&\nn\\
   \br{\rm x}^3(\tau)\kt\!\aeq\!\br{\rm x}^3\kt +
   \tau \br\dot{\rm x}^3\kt = \br{\rm x}^3\kt + \tau \br\frac{{\rm p}_3}{H'}\kt,
   \label{f-mat-x3}\\ &&\nn\\
   \br\dot{\rm x}^3\kt :=  {\cal E} \!\!\!\!\!&&\!\!\!\!\!
   \left\{\,\sum_{n=0}^{\infty}\sum_{\ell=-\infty}^{\infty}
   \frac{n!}{(n+|\ell|)!}\,\,s^{2n+|\ell|} u^{|\ell|}\!
   \left(L_n^{\!|\ell|}(u)\right)^2\!\!\!\int_{-\infty}^{+\infty}\!\!\!
   \!\!\frac{k_3\,e^{-\frac{2}{\lambda_3^2}(k_3-\br{\rm p}_3\kt)^2}\,d k_3}
   {\sqrt{\Lambda(2n+1-\ell+|\ell|)+1+k_3^2}}\right\} ,~~~~~~~~~~
   \label{f-mat-vel}\\
   &&\nn\\
   \br{\rm p}_1(\tau)\kt\!\aeq\!\br{\rm p}_1\kt +
   \frac{\Lambda}{2}\,\br{\rm x}^2(\tau)\kt, ~~~~~~~~~~
   \br{\rm p}_2(\tau)\kt = -\,\frac{\Lambda}{2}\,
   \br{\rm x}^1(\tau)\kt, ~~~~~~~~~~
   \br{\rm p}_3(\tau)\kt = \br{\rm p}_3\kt,
    \label{f-mat-p123}
   \eea}
where
    \bea
    &&\lambda_\perp := \frac{\sqrt{2\hbar k_\perp}}{mc}, ~~~~~~
    \lambda_3: = \frac{\sqrt{2\hbar k_3}}{mc}, ~~~~~~
    \Lambda: = \frac{e\hbar B}{(mc)^2}, ~~~~~~
    s := \frac{\Lambda - \lambda_\perp^2}{\Lambda + \lambda_\perp^2}, ~~~~~~
    u := \frac{2 \Lambda \br{\rm p}_1\kt^2}{
    \Lambda^2 - \lambda_\perp^4},~~~~~\nn\\
    &&{\cal F} := \sqrt{\frac{2}{\pi}}\,
    \frac{8\lambda_\perp^2 \Lambda \br{\rm p}_1\kt}{
   \lambda_3(\Lambda + \lambda_\perp^2)^3}
   \,e^{-\frac{2 \br{\rm p}_1\kt^2}{(\Lambda + \lambda_\perp^2)}}, ~~~~~~~~~
   {\cal E} := \sqrt{\frac{2}{\pi}}\,\frac{4\lambda_\perp^2 \Lambda}{
   \lambda_3(\Lambda + \lambda_\perp^2)^2}
   \,e^{-\frac{2 \br{\rm p}_1\kt^2}{(\Lambda + \lambda_\perp^2)}},
    \nn\\
    &&\Theta_{n\ell}(q) := \tau\left(\sqrt{\Lambda(2n+2\ell+3)+1+q^2}-
   \sqrt{\Lambda(2n+2\ell+1)+1+q^2}\right),\nn
   \eea
$\br{\rm p}_i\kt$ with $i=1,2,3$ are the initial
(kinetic) momentum expectation value in
$x^i$-direction\footnote{Note that because we consider $\br{\rm
x}_i\kt = 0, i=1,2,3$, in view of Eq.~(\ref{pken1}), $\br{\rm
p}_i\kt$ are the kinetic momentum of the initial coherent state.},
and $\tau:=x^0/\lambda_c = (m c^2)t/\hbar$ is the dimensionless
time parameter. It is instructive to note that, for example, for
$\pi^+$ meson, $\Lambda\approx 2.8\times 10^{-15}$($B$/Teslas).

Employing the ``Gauss-Hermite routine of integration"
\cite{numrec}, we have written and used a computer code in C++ to
numerically perform the integrals and sums appearing in the above
and following expressions for various expectation values. We
present a summary of the results in Figs.~\ref{helix} --
\ref{x1x2-nr} which we briefly elude to below.

Fig.~\ref{helix} shows two typical trajectories traced by the
expectation value of the position operator for the magnetic field
parameter $\Lambda=0.001$, momentum expectation value
$\br\Pi\kt=2, \br{\rm p}_3\kt=1.6$, and two cases of the initial
widths of coherent state: $\lambda_3=10^{-3},
\lambda_\perp\simeq\sqrt{\Lambda}$ and
$\lambda_3=\lambda_\perp=0.25$. As indicated in this figure, for
the first values of $\lambda_3$ and $\lambda_\perp$ the helix is
very close to the classical result.\footnote{In all of our
numerical results a small fraction of the deviation from the
classical result is due to the errors in numerical calculations.}

Figs.~\ref{x1x2-1} -- \ref{x1x2-4} show $\br{\rm x}^1(\tau)\kt$
and $\br{\rm x}^2(\tau)\kt+1$ as a function of time for five
classical periods of precession. $\br{\rm x}^1(\tau)\kt$ and
$\br{\rm x}^2(\tau)\kt+1$ oscillate respectively like damped sine
and cosine functions. For $\lambda_\perp\rightarrow\sqrt{\Lambda}$
and smaller values of $\lambda_3$ the curves are closer to the
corresponding classical curves. The effects of the width
$\lambda_3$ is greater than the effect of $\lambda_\perp$. Also,
changing the initial transverse momentum does not affect the
behavior of $\br{\rm x}^1(\tau)\kt$ and $\br{\rm x}^2(\tau)\kt$,
though their magnitude clearly depends on the initial transverse
momentum. The higher the initial momentum and the smaller the
magnetic field become the closer the curves of $\br{\rm
x}^1(\tau)\kt$ and $\br{\rm x}^2(\tau)\kt+1$ get to the
corresponding classical curves.

Fig.~\ref{p1p2} shows $\br{\rm p}_1(\tau)\kt-1$, $\br{\rm
p}_2(\tau)\kt$ and expectation value of transverse kinetic momenta
$\br\Pi_1(\tau)\kt$ and $\br\Pi_2(\tau)\kt$ as functions of time
for five classical periods of precession. $\br{\rm
p}_1(\tau)\kt-1$ and $\br\Pi_1(\tau)\kt$ (respectively
$\br\Pi_2(\tau)\kt$ and $\br{\rm p}_2(\tau)\kt$) oscillate like a
damped cosine (resp.\ sine) function. As
$\lambda_\perp\rightarrow\sqrt{\Lambda}$ and $\lambda_3\to 0$, the
curves tend to their corresponding classical curves. Since the
behavior of the momentum expectation value is similar to the
position expectation value, we can conclude that as we increase
the initial momentum and decrease the magnetic field the graphs of
the components of the momentum expectation value tend to those of
the classical momenta.

As indicated in these figures, the trajectories traced by the
expectation value of the position and momentum operators do not
coincide with the classical trajectories. The position expectation
values trace a helix with a constant gyration center and a
decreasing radius. Moreover the period of precession is smaller
than the classical period. Note that because $\sqrt{(\br{\rm
x}^1\kt-\br{\rm x}^1_{\rm g.c.}\kt)^2+(\br{\rm x}^2\kt-\br{\rm
x}^2_{\rm g.c.}\kt)^2}$ is not the gyration radius, a decrease in
this quantity as depicted in the graphs of Figs.~\ref{x1x2-1} --
\ref{x1x2-4} does not mean that the expectation value of the
radius operator decreases. Indeed, if we identify the operators of
the gyration center with
    \be
    {\rm x}^1_{\rm g.c.} := \frac{{\rm x}^1}{2} +
    \frac{{\rm p}_2}{e B} , ~~~~~~~~~~~
    {\rm x}^2_{\rm g.c.} := \frac{{\rm x}^2}{2} -
    \frac{{\rm p}_1}{e B},
    \label{gyr-cent-op}
    \ee
which we have obtained by quantization of their classical
counterparts, and use (\ref{f-mat-x1}) -- (\ref{f-mat-p123}), we
can easily see that the expectation value of these operators in
the coherent state vector $|\vec\zeta\kt$ are constant. This means
that the gyration center is a constant point. Similarly we obtain
the energy, radius, and angular momentum operators by quantization
the corresponding classical quantities (\ref{mf-ch}),
(\ref{gyr-rad}) and (\ref{angm}). The expectation value of these
operators in the coherent state vector $|\vec\zeta\kt$ are also
time-independent. They are given by
   {\small \bea
   \br E\kt \!\aeq\! {\cal E}
   \left\{\,\sum_{n=0}^{\infty}\sum_{\ell=-\infty}^{\infty}
   \frac{n!}{(n+|\ell|)!}\,\,s^{2n+|\ell|} u^{|\ell|}
   \!\left(L_n^{\!|\ell|}(u)\right)^2 \right.
   \nn\\
   &&\vspace{2cm}\left.~~~~~~~~~~~~\times\int_{-\infty}^{+\infty}
   \!\!\!\!\!\!d k_3\,e^{-\frac{2}{\lambda_3^2}(k_3-\br{\rm p}_3\kt)^2}
   \!\!\sqrt{\Lambda(2n+1-\ell+|\ell|)+1+k_3^2}\, \right\} ,
   \label{E-magexp}\\
   \br L_3\kt \!\aeq\! \frac{4\Lambda\lambda_\perp^2
   \,e^{-\frac{2 \br{\rm p}_1\kt^2}{(\Lambda + \lambda_\perp^2)}}}
   {(\Lambda + \lambda_\perp^2)^2}
   \left\{\,\sum_{n=0}^{\infty}\sum_{\ell=-\infty}^{\infty}\!\!\ell\,
   \,\frac{n!}{(n+|\ell|)!}\,\,s^{2n+|\ell|} u^{|\ell|}
   \!\left(L_n^{\!|\ell|}(u)\right)^2 \right\} = 0 ,
   \label{ang-expect}
   \\
   \br R\kt \!\aeq\! \frac{4\lambda_\perp^2 \sqrt{\Lambda}
   \,e^{-\frac{2 \br{\rm p}_1\kt^2}{(\Lambda + \lambda_\perp^2)}}}
   {(\Lambda + \lambda_\perp^2)^2}
   \left\{\,\sum_{n=0}^{\infty}\sum_{\ell=-\infty}^{\infty}
   \frac{n!}{(n+|\ell|)!}\,\,s^{2n+|\ell|} u^{|\ell|}
   \!\left(L_n^{\!|\ell|}(u)\right)^2\!\sqrt{2n+1-\ell+|\ell|}\right\} ,
   \label{R-magexp} \\
   \br R^2\kt \!\aeq\! \frac{4\lambda_\perp^2
   \,e^{-\frac{2 \br{\rm p}_1\kt^2}{(\Lambda + \lambda_\perp^2)}}}
   {(\Lambda + \lambda_\perp^2)^2}
   \left\{\,\sum_{n=0}^{\infty}\sum_{\ell=-\infty}^{\infty}
   \frac{n!}{(n+|\ell|)!}\,\,s^{2n+|\ell|} u^{|\ell|}
   \!\left(L_n^{\!|\ell|}(u)\right)^2 \left(2n+1-\ell+|\ell|\right)\right\}
   \nn\\
   \!\aeq\! \frac{\br{\rm p}_1^2\kt}{\Lambda^2} +
   \frac{(\Lambda^2 + \lambda_\perp^4)}{2\Lambda^2\lambda_\perp^2}
   = \br R_{\rm g.c.}^2\kt\,,
   \label{R2-expect}
   \eea
where we have performed the summations appearing in
(\ref{R2-expect}) by means of some useful identities listed in
\cite{Ryzh,Lebedev}, and in the last equality we have made use of
(\ref{angm}) and (\ref{ang-expect}). Note that $\br R\kt$, $\br
R^2\kt$ and $\br L_3\kt$ do not depend on the width $\lambda_3$.

Tables~\ref{tab1} and \ref{tab2} show some typical values of the
expectation values of energy, $x^3$-component of the velocity,
radius, square of radius, and dispersion in radius. The deviation
from the classical values which is generally small diminishes for
larger values of the initial momentum and the smaller values of
the magnetic field.

\begin{table}[p]
\begin{center}\begin{tabular}{||c|c||c|c|c|c|c||}
\hline\hline
 \multicolumn{2}{||c||}{\!\!$\br\Pi\kt,~\br{\rm p}_1\kt,~\lambda_\perp$ and $\lambda_3$\!\!}&
 \multicolumn{1}{c|}{$\br E\kt/E_{\rm cl.}$}&
 \multicolumn{1}{c|}{$\br\dot{\rm x}^3\kt/\dot{x}^3_{\rm cl.}$}&
  \multicolumn{1}{c|}{$\br R\kt/R_{\rm cl.}$}&\multicolumn{1}{c|}{$\br R^2\kt/R_{\rm cl.}^2$}&
  \multicolumn{1}{c||}{$\Delta R/R_{\rm cl.}$} \\
\hline\hline
  &  \multicolumn{1}{c||}{$\lambda_3=10^{-3},~\lambda_\perp\simeq\sqrt{\Lambda}$} &
  \multicolumn{1}{c|}{$1.00086$} &  \multicolumn{1}{c|}{$0.99943$} &
  \multicolumn{1}{c|}{$1.00174$} &  \multicolumn{1}{c|}{$1.00694$} &
  \multicolumn{1}{c||}{$0.05887$} \\
 \cline{2-2} \cline{3-3} \cline{4-4} \cline{5-5} \cline{6-6} \cline{7-7}
 $\br\Pi\kt=2$ &  \multicolumn{1}{c||}{$\lambda_3=0.5,~\lambda_\perp\simeq\sqrt{\Lambda}$} &
 \multicolumn{1}{c|}{$1.00394$} &  \multicolumn{1}{c|}{$0.99022$} &
 \multicolumn{1}{c|}{$1.00174$} &  \multicolumn{1}{c|}{$1.00694$} &
 \multicolumn{1}{c||}{$0.05887$} \\
 \cline{2-2} \cline{3-3} \cline{4-4} \cline{5-5} \cline{6-6} \cline{7-7}
 $\br{\rm p}_1\kt=1.2$ &  \multicolumn{1}{c||}{$\lambda_3=10^{-3},~\lambda_\perp=0.5$} &
 \multicolumn{1}{c|}{$1.01067$} &  \multicolumn{1}{c|}{$0.99286$} &
 \multicolumn{1}{c|}{$1.02199$} &  \multicolumn{1}{c|}{$1.08694$} &
 \multicolumn{1}{c||}{$0.20611$} \\
 \cline{2-2} \cline{3-3} \cline{4-4} \cline{5-5} \cline{6-6} \cline{7-7}
  &  $\lambda_3=\lambda_\perp=0.5$ &  $1.01375$ &  $0.98383$ &  $1.02199$
  &  $1.08694$ & $0.20611$ \\
\hline\hline
  &  \multicolumn{1}{c||}{$\lambda_3=10^{-3},~\lambda_\perp\simeq\sqrt{\Lambda}$} &
  \multicolumn{1}{c|}{$1.00074$} &  \multicolumn{1}{c|}{$0.99976$} &
  \multicolumn{1}{c|}{$1.00097$} &  \multicolumn{1}{c|}{$1.00391$} &
  \multicolumn{1}{c||}{$0.04417$} \\
 \cline{2-2} \cline{3-3} \cline{4-4} \cline{5-5} \cline{6-6} \cline{7-7}
 $\br\Pi\kt=2$ &  \multicolumn{1}{c||}{$\lambda_3=0.5,~\lambda_\perp\simeq\sqrt{\Lambda}$} &
 \multicolumn{1}{c|}{$1.00521$} &  \multicolumn{1}{c|}{$0.98663$} &
 \multicolumn{1}{c|}{$1.00097$} &  \multicolumn{1}{c|}{$1.00391$} &
 \multicolumn{1}{c||}{$0.04417$} \\
 \cline{2-2} \cline{3-3} \cline{4-4} \cline{5-5} \cline{6-6} \cline{7-7}
 $\br{\rm p}_1\kt=1.6$ &  \multicolumn{1}{c||}{$\lambda_3=10^{-3},~\lambda_\perp=0.5$} &
 \multicolumn{1}{c|}{$1.00932$} &  \multicolumn{1}{c|}{$0.99691$} &
 \multicolumn{1}{c|}{$1.01230$} &  \multicolumn{1}{c|}{$1.04891$} &
 \multicolumn{1}{c||}{$0.15539$} \\
 \cline{2-2} \cline{3-3} \cline{4-4} \cline{5-5} \cline{6-6} \cline{7-7}
  &  \multicolumn{1}{c||}{$\lambda_3=\lambda_\perp=0.5$} &
  \multicolumn{1}{c|}{$1.01375$} &  \multicolumn{1}{c|}{$0.98383$} &
  \multicolumn{1}{c|}{$1.01230$} &  \multicolumn{1}{c|}{$1.04891$} &
  \multicolumn{1}{c||}{$0.15539$} \\
\hline\hline
  &  \multicolumn{1}{c||}{$\lambda_3=\lambda_\perp=0.25$} &
  \multicolumn{1}{c|}{$1.00063$} &  \multicolumn{1}{c|}{$0.99936$} &
  \multicolumn{1}{c|}{$1.00089$} &  \multicolumn{1}{c|}{$1.00356$} &
  \multicolumn{1}{c||}{$0.04218$} \\
 \cline{2-2} \cline{3-3} \cline{4-4} \cline{5-5} \cline{6-6} \cline{7-7}
 $\br\Pi\kt=5$ &  \multicolumn{1}{c||}{$\lambda_3=\lambda_\perp=0.5$} &
 \multicolumn{1}{c|}{$1.00245$} &  \multicolumn{1}{l|}{$0.99745$} &
 \multicolumn{1}{c|}{$1.00348$} &  \multicolumn{1}{c|}{$1.01391$} &
 \multicolumn{1}{c||}{$0.08325$} \\
 \cline{2-2} \cline{3-3} \cline{4-4} \cline{5-5} \cline{6-6} \cline{7-7}
 $\br{\rm p}_1\kt=3$ &  \multicolumn{1}{c||}{$\lambda_3=0.25,~\lambda_\perp=0.5$} &
 \multicolumn{1}{c|}{$1.00211$} &  \multicolumn{1}{c|}{$0.99849$} &
 \multicolumn{1}{c|}{$1.00348$} &  \multicolumn{1}{c|}{$1.01391$} &
 \multicolumn{1}{c||}{$0.08325$} \\
 \cline{2-2} \cline{3-3} \cline{4-4} \cline{5-5} \cline{6-6} \cline{7-7}
  &  $\lambda_3=0.5,~\lambda_\perp=0.25$ &  $1.00097$ &  $0.99831$
  &  $1.00089$ & $1.00356$ & $0.04218$ \\
\hline\hline
\end{tabular}\end{center}
\centerline{
    \parbox{14cm}{
    \caption{Expectation values of energy,
    $x^3$-component of the velocity, radius, square of radius, and
    dispersion in radius are given for magnetic field parameter
    $\Lambda=0.01$ and various initial momenta and widths of the
    coherent state. The coherent states with smaller width and
    larger momentum display more pronounced classical behavior.}
    \label{tab1}}}
\vspace{2cm}
%\end{table}
%\begin{table}[h]
\begin{center}\begin{tabular}{||c|c||c|c|c|c|c||}
\hline\hline
 $\lambda_\perp$ and $\lambda_3$ & $\Lambda$ & $\br E\kt/E_{\rm cl.}$ & $\br\dot{\rm x}^3\kt/\dot{x}^3_{\rm cl.}$
 & $\br R\kt/R_{\rm cl.}$ & $\br R^2\kt/R_{\rm cl.}^2$ & $\Delta R/R_{\rm cl.}$ \\
 \hline\hline
  $\lambda_\perp=0.25$ & \multicolumn{1}{c||}{$\Lambda=0.1$} &
  \multicolumn{1}{c|}{$1.01029$} &  \multicolumn{1}{c|}{$0.99133$} &  \multicolumn{1}{c|}{$1.01952$} &
  \multicolumn{1}{c|}{$1.07725$} & \multicolumn{1}{c||}{$0.19453$} \\
 \cline{2-2} \cline{3-3} \cline{4-4} \cline{5-5} \cline{6-6} \cline{7-7}
 $\lambda_3=0.25$  &  \multicolumn{1}{c||}{$\Lambda=10^{-4}$} &
 \multicolumn{1}{c|}{$1.00344$} &  \multicolumn{1}{c|}{$0.99594$} &  \multicolumn{1}{c|}{$1.00544$} &
  \multicolumn{1}{c|}{$1.02171$} & \multicolumn{1}{c||}{$0.10385$} \\
 \hline\hline
  $\lambda_\perp=0.5$ & \multicolumn{1}{c||}{$\Lambda=0.1$} &
  \multicolumn{1}{c|}{$1.01547$} &  \multicolumn{1}{c|}{$0.98264$} &
  \multicolumn{1}{c|}{$1.02553$} & \multicolumn{1}{c|}{$1.10069$} &
  \multicolumn{1}{c||}{$0.22128$} \\
 \cline{2-2} \cline{3-3} \cline{4-4} \cline{5-5} \cline{6-6} \cline{7-7}
 $\lambda_3=0.5$ &  \multicolumn{1}{c||}{$\Lambda=10^{-4}$} &
 \multicolumn{1}{c|}{$1.01372$} &  \multicolumn{1}{c|}{$0.98385$} &
 \multicolumn{1}{c|}{$1.02195$} & \multicolumn{1}{c|}{$1.08681$} &
 \multicolumn{1}{c||}{$0.20595$} \\
 \hline\hline
\end{tabular}\end{center}
\centerline{
    \parbox{14cm}{
    \caption{Expectation values of energy,
    $x^3$-component of the velocity, radius, square of radius, and
    dispersion in radius are given for magnetic field parameters
    $\Lambda=0.1,~10^{-4}$,
    initial momenta $\br\Pi\kt=2,~\br{\rm p}_3\kt=1.2$, and various
    widths $\lambda_\perp$ and $\lambda_3$.
    For smaller values of the magnetic field the results are
    closer to the corresponding classical quantities.}
    \label{tab2}}}
    \end{table}

In the limit $B\rightarrow 0$ (equivalently $\Lambda\rightarrow
0$), Eqs.~(\ref{f-mat-x1}) -- (\ref{f-mat-vel}) and
(\ref{E-magexp}) reduce to the corresponding equations for a free
KG field. For the case that the expectation value of the initial
transverse momentum vanishes, i.e., $\br{\rm p}_i\kt=0$ for
$i=1,2$, we find\footnote{These equations are to be compared with
(\ref{1xdot11}), (\ref{1xtexp11}) and (\ref{1energy2}).}
    \bea
    \br{\rm x}^3(\tau)\kt \aeq
    \frac{\tau}{\lambda_3} \sqrt{\frac{2}{\pi}}
    \int_{-\infty}^{+\infty} d k_3\,\frac{k_3\,
    e^{-\frac{2}{\lambda_3^2} (k_3 - \br{\rm p}_3\kt)^2}}{
    \sqrt{1+k_3^2}} , ~~~~~~~
    \br{\rm x}^1(\tau)\kt= \br{\rm x}^2(\tau)\kt= 0, \nn\\
    \br E\kt \aeq \frac{1}{\lambda_3} \sqrt{\frac{2}{\pi}}
    \int_{-\infty}^{+\infty} d k_3\,\sqrt{1+k_3^2}\,
    e^{-\frac{2}{\lambda_3^2} (k_3 - \br{\rm p}_3\kt)^2}.\nn
    \eea
Hence we have a free motion along $x^3$-direction.

Next, we determine the uncertainty relationship for ${\rm x}^3$
and ${\rm p}_3$ at an arbitrary time $\tau\in\R$.\footnote{The
calculation of the uncertainty relationship for ${\rm x}^i$ and
${\rm p}_i$, with $i=1,2$ is much more complicated, and we were
not able to simplify them to a presentable form.} This yields
   \be
   (\Delta {\rm x}^3(\tau))(\Delta {\rm p}_3(\tau))=\frac{1}{2}
   \sqrt{1 + \lambda_3^2 (\Delta\dot{\rm x}^3)^2\,\tau^2},
   \label{uncer-x3-p3}
   \ee
where $(\Delta\dot{\rm x}^3)^2:= \br (\dot{\rm x}^3)^2\kt- \br
\dot{\rm x}^3\kt^2$, $\br \dot{\rm x}^3\kt$ is given by
(\ref{f-mat-vel}), and
   \bea
   \br (\dot{\rm x}^3)^2\kt \aeq {\cal E}
   \left\{\,\sum_{n=0}^{\infty}\sum_{\ell=-\infty}^{\infty}
   \frac{n!}{(n+|\ell|)!}\,\,s^{2n+|\ell|} u^{|\ell|}
   \!\left(L_n^{\!|\ell|}(u)\right)^2\!\!\!\int_{-\infty}^{+\infty}
   \!\!\!\frac{k_3^2\,
   e^{-\frac{2}{\lambda_3^2}(k_3-\br{\rm p}_3\kt)^2}\,d k_3}
   {\Lambda(2n+1-\ell+|\ell|)+1+k_3^2}\right\} .
   \nn
   \eea
It is not difficult to check that in the limit $B\rightarrow 0$
(equivalently $\Lambda\rightarrow 0$) Eq.~(\ref{uncer-x3-p3})
reproduces the uncertainty relationship (\ref{tunc2}) for the free
field.

Fig.~\ref{magunc} shows a plot of the right-hand side of
(\ref{uncer-x3-p3}) for different values of the initial momentum
and the magnetic field. For the range of values of $\Lambda$ that
are used in the graphs, $(\Delta {\rm x}^3(\tau))(\Delta {\rm
p}_3(\tau))$ is an increasing function of the magnetic field. This
also implies, in view of (\ref{uncer-x3-p3}), that the presence of
the magnetic field enhances the spreading of the wave packet.
Moreover, similarly to the case of a free KG field, faster moving
wave packets have a lower spreading rate.

Finally, using Eqs.~(\ref{exp-cph}), (\ref{mg-dop}),
(\ref{3x-coh}), (\ref{eigenfunc}), (\ref{eigenval}), and inserting
(\ref{simpch}) in (\ref{3x-coh}), we derive the functional form of
a coherent KG field:
    {\small\bea
    \psi_a^{(\vec\zeta)}(x):=\psi_a^{(+,\vec\zeta_+)}(x)
    \!\aeq\!\frac{\sqrt{\Lambda {\cal E}}\,e^{-v/2}}{2\pi\sqrt{\kappa (1+a)}}
    \left\{\,\sum_{n=0}^{\infty}\sum_{\ell=-\infty}^{\infty}\!\!\!
    \left(\frac{n!}{(n+|\ell|)!}\right)\!e^{i\ell\varphi}
    (-s)^{n+|\ell|/2} (u v)^{|\ell|/2} L_n^{\!|\ell|}(v)
    L_n^{\!|\ell|}(u) \right.
    \nn\\
    &&\vspace{2cm}\left.\times\int_{-\infty}^{+\infty}
    \!\!\!\!\!\!\!d k_3\,e^{-\frac{1}{\lambda_3^2}(k_3-\br{\rm p}_3\kt)^2}
    e^{i k_3 x^3}
    \frac{e^{-i (\tau-\tau_0) \sqrt{\Lambda(2n+1-\ell+|\ell|)+1+k_3^2}}}
    {\left[\Lambda(2n+1-\ell+|\ell|)+1+k_3^2\right]^{1/4}}\right\},
    \label{kgf-mag}
    \eea}
\noindent where $v\!:=\!\Lambda\rho^2/2$. In the nonrelativistic
limit ($c\rightarrow\infty$), $\psi_a^{(\vec\zeta)}(x)$ tends to
the nonrelativistic coherent wave function (\ref{3x-coh}). This is
consistent with Fig.~\ref{kgx3rho-pvar} which gives the plots of
$|\fc(x^0_0,\vec x)|^2$ and $|\psi_a^{(\vec\zeta)}(x^0_0,\vec
x)|^2$ for various values of the momentum expectation value $\br
p\kt_\epsilon=k\beta_\epsilon$ and the dimensionless widths
$\lambda_\perp, \lambda_3$. As seen from this figure,
$|\psi_a^{(\vec\zeta)}(x^0_0,\vec x)|^2$ depends on the
expectation value of the initial momentum. In particular for
smaller values of the initial momentum it tends to the
nonrelativistic probability density $|\fc(x^0_0,\vec x)|^2$.

Fig.~\ref{kgx3rho-mvar} shows the plots of
$|\psi_a^{(\vec\zeta)}(x^0_0,\vec x)|^2$ for three different
values of the magnetic field parameter $\Lambda$. A surprising
behavior depicted in these plots, which is not evident from
(\ref{kgf-mag}), is that $|\psi_a^{(\vec\zeta)}(x^0_0,\vec x)|^2$
does not depend on the magnetic field. This observation suggests
identifying the coherent states $|\vec\zeta\kt$ of
Eq.~(\ref{3d-coh}) as the appropriate nonrelativistic coherent
states of a charged particle in a homogeneous magnetic field. To
the best of our knowledge this identification has not been
previously considered in the literatures \cite{kowalski, malkin,
cohen}.

Adopting (\ref{3d-coh}) as the defining relation for
nonrelativistic coherent states, we have computed the expectation
values of position, momentum, energy, $x^3$-component of the
velocity, radius, and angular momentum operators using the same
method as for the relativistic coherent states. The result is
   \bea
   \br{\rm x}^1(\tau)\kt_{\rm nr} \aeq \frac{\br{\rm p}_1\kt}{\Lambda}\,
   \sin{\Lambda\tau}, ~~~~~~
   \br{\rm x}^2(\tau)\kt_{\rm nr} = \frac{\br{\rm p}_1\kt}{\Lambda}\,
   \left(\cos{\Lambda\tau} - 1\right), ~~~~~~
   \br{\rm x}^3(\tau)\kt_{\rm nr} = \br{\rm x}^3\kt +
   \br{\rm p}_3\kt\,\tau,~~~~~~~~~~~
   \label{x123-nr}\\
   \br{\rm p}_1(\tau)\kt_{\rm nr} \aeq \frac{\br{\rm p}_1\kt}{2}\,
   \left(\cos{\Lambda\tau} + 1\right), ~~~~~~~~
   \br{\rm p}_2(\tau)\kt_{\rm nr} = -\,\frac{\br{\rm p}_1\kt}{2}\,
   \sin{\Lambda\tau}, ~~~~~~~~
   \br{\rm p}_3(\tau)\kt_{\rm nr} = \br{\rm p}_3\kt.
   \label{p123-nr}\\
   \br E\kt_{\rm nr} \aeq 1 + \frac{\lambda_3^2}{8} + \frac{\br{\rm
   p}_3\kt^2}{2} + \frac{\Lambda^2 \br R^2\kt_{\rm nr}}{2}
   =1+\frac{\lambda_3^2}{8}+\frac{\br{\rm
   p}_3\kt^2}{2}+\frac{\br\Pi_\perp^2\kt_{\rm nr}}{2} ,~~~~~~~
   \br\dot{\rm x}^3(\tau)\kt_{\rm nr}=\br{\rm p}_3\kt ,
   \label{eng-nr}
   \eea
and $\br R\kt_{\rm nr}$, $\br R^2\kt_{\rm nr}$ and $\br
L_3\kt_{\rm nr}$ have the same form as relativistic case. For
$B=0$ (or $\Lambda=0$) Eq.~(\ref{eng-nr}) tends to the well-known
result for the free particle (compare with (\ref{nonrelenrg})).
These nonrelativistic expectation values do not depend on the
widths $\lambda_\perp$ and $\lambda_3$ and coincide with the
corresponding classical quantities. Using our numerical method, we
have compared the relativistic and nonrelativistic results and
checked that the former reproduces the latter, namely
(\ref{x123-nr}) -- (\ref{eng-nr}), in the nonrelativistic limit.
Fig.~\ref{x1x2-nr} provides a graphical demonstration of this
comparison.

Table \ref{tab3} shows the energy and velocity expectation values
obtained using the relativistic expression for a small value of
the initial momentum. It also includes the nonrelativistic results
obtained using (\ref{eng-nr}). Relativistic and nonrelativistic
calculations yield the same values for the expectation value of
the radius $R$, square of radius $R^2$, and angular momentum
$L_3$.
\begin{table}[h]
\begin{center}\begin{tabular}{||c|c||c|c|c||}
\hline\hline
 $\lambda_\perp$ and $\lambda_3$ & $\Lambda$ & $\br E\kt/E_{\rm cl.}$ & $\br E\kt_{\rm nr}/E_{\rm cl.}$
 & $\br\dot{\rm x}^3\kt/\dot{x}^3_{\rm cl.}$ \\
 \hline\hline
  $\lambda_\perp=0.25$ & \multicolumn{1}{c||}{$\Lambda=0.1$} &
  \multicolumn{1}{c|}{$1.06127$} &  \multicolumn{1}{c|}{$1.06344$} &  \multicolumn{1}{c||}{$0.93013$}
   \\
 \cline{2-2} \cline{3-3} \cline{4-4} \cline{5-5}
 $\lambda_3=0.25$  &  \multicolumn{1}{c||}{$\Lambda=10^{-4}$} &
 \multicolumn{1}{c|}{$1.02300$} &  \multicolumn{1}{c|}{$1.02344$} &  \multicolumn{1}{c||}{$0.96381$}
  \\
 \hline\hline
  $\lambda_\perp=0.5$ & \multicolumn{1}{c||}{$\Lambda=0.1$} &
  \multicolumn{1}{c|}{$1.09724$} &
  \multicolumn{1}{c|}{$1.10375$} & \multicolumn{1}{c||}{$0.87193$}
  \\
 \cline{2-2} \cline{3-3} \cline{4-4} \cline{5-5}
 $\lambda_3=0.5$ &  \multicolumn{1}{c||}{$\Lambda=10^{-4}$} &
 \multicolumn{1}{c|}{$1.08764$} &
 \multicolumn{1}{c|}{$1.09375$} & \multicolumn{1}{c||}{$0.87959$}
 \\
 \hline\hline
\end{tabular}\end{center}
\centerline{
    \parbox{14cm}{
    \caption{Energy and velocity expectation values
    are given for the initial momentum expectation value
    $\br\Pi\kt=0.001$, magnetic field parameters
    $\Lambda=0.1,~10^{-4}$, and various widths
    $\lambda_\perp$ and $\lambda_3$. For all values of these
    parameters $\br\dot{\rm x}^3\kt_{\rm nr}/\dot{x}^3_{\rm cl.}=1$.
    Hence the data confirms that our relativistic
    calculations have the correct nonrelativistic limit.}
    \label{tab3}}}
    \end{table}

\section{Conclusion}

In \cite{p59} we give a formulation of the quantum mechanics of
first quantized scalar fields which is based on the construction
of a genuine Hilbert space. This is determined by a one-parameter
family of inner products $(\cdot,\cdot)_a$ where $a\in(-1,1)$. The
Hilbert spaces ${\cal H}_a$ associated with different allowed
values of $a$ are unitary-equivalent to $L^2(\R^3)\oplus
L^2(\R^3)$. This allows for a straightforward construction of an
appropriate pair of relativistic position and momentum operators
and the corresponding relativistic coherent states.

In this paper we offer an explicit construction and a detailed
investigation of the coherent states for both charged and neutral
KG fields that are either free or interact with a constant
homogeneous magnetic field. Our strategy is to construct coherent
states in the two-component Foldy representation and pull them
back using the appropriate unitary transformation to obtain
coherent KG fields. In contrast to the earlier approaches to this
problem, ours is free from the problems associated with the
charge-superselection rule.

The general behavior of our coherent states are similar to that of
a classical particle in both free and interacting cases. Moreover,
in the nonrelativistic limit our results coincide with those of
nonrelativistic quantum mechanics.

%\renewcommand{\baselinestretch}{1}
%\newpage

\newpage
\begin{figure}
\vspace{-3cm}
\centerline{\includegraphics[width=.5\columnwidth]{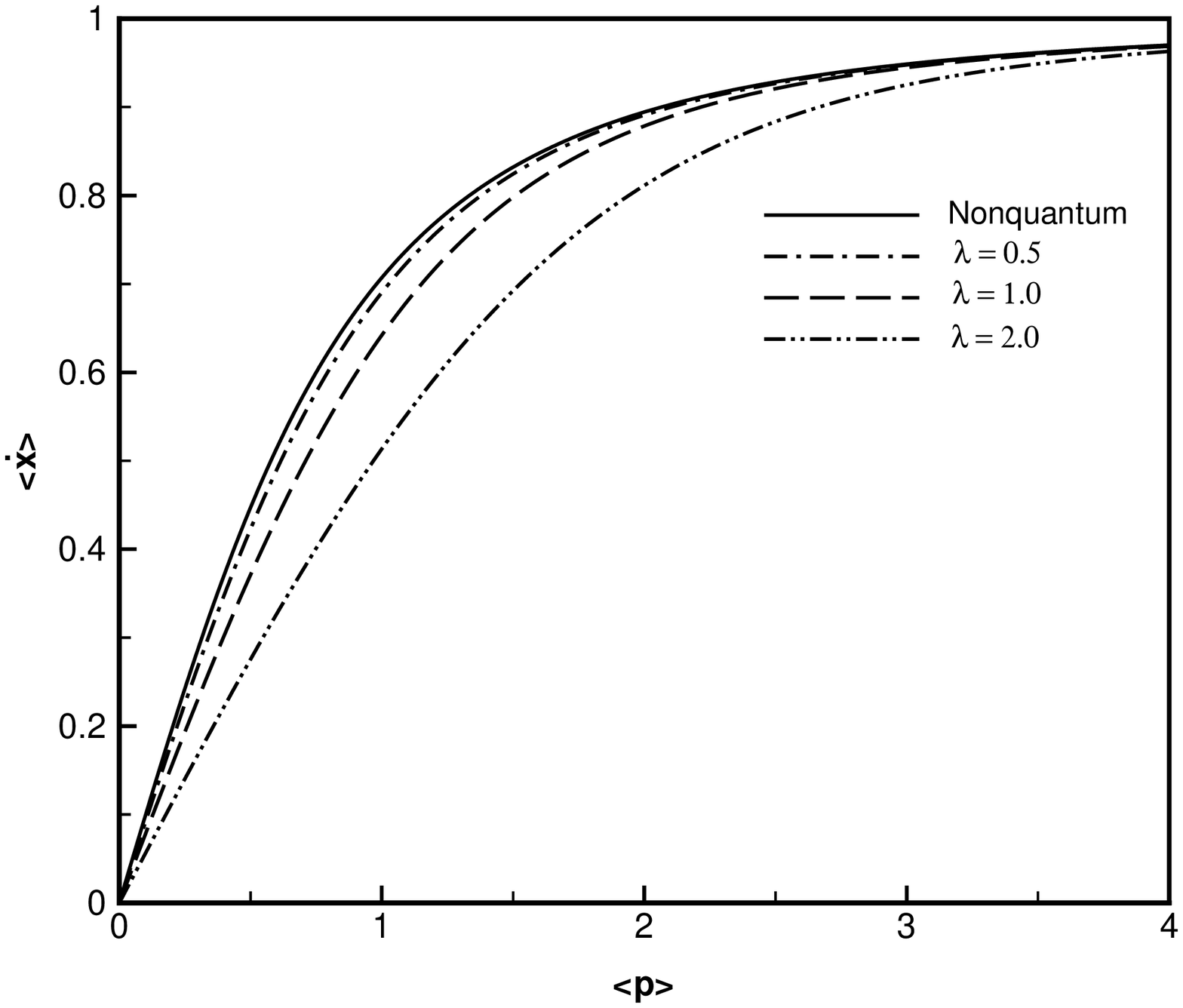}
\includegraphics[width=.5\columnwidth]{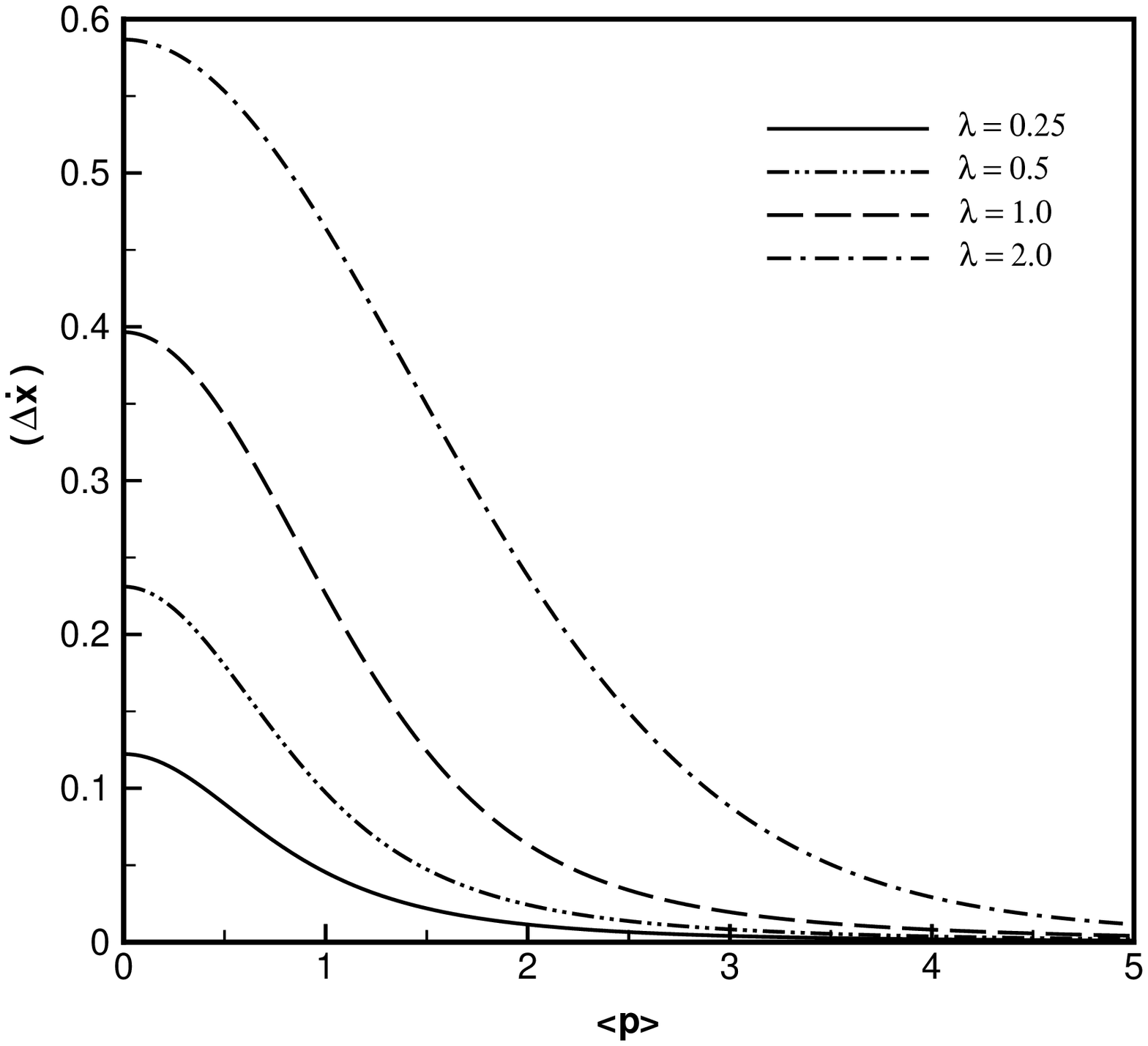}}
\caption{Graphs of the velocity expectation value $\br\dot{\rm
x}\kt$ (left) and the dispersion $(\Delta\dot{\rm x})$ (right) as
functions of the momentum expectation value $\br{\rm p}\kt$ for
coherent states of a free particle with different values of
$\lambda$ and the corresponding classical (nonquantum) curve:
Momentum and velocity are given in units of $m c$ and $c$,
respectively. For small values of $\lambda$ the quantum curves
tend to the classical curve. The dispersion in velocity decreases
as the velocity increases.} \label{xdotfig}
\end{figure}

\begin{figure}
\centerline{\includegraphics[width=.5\columnwidth]{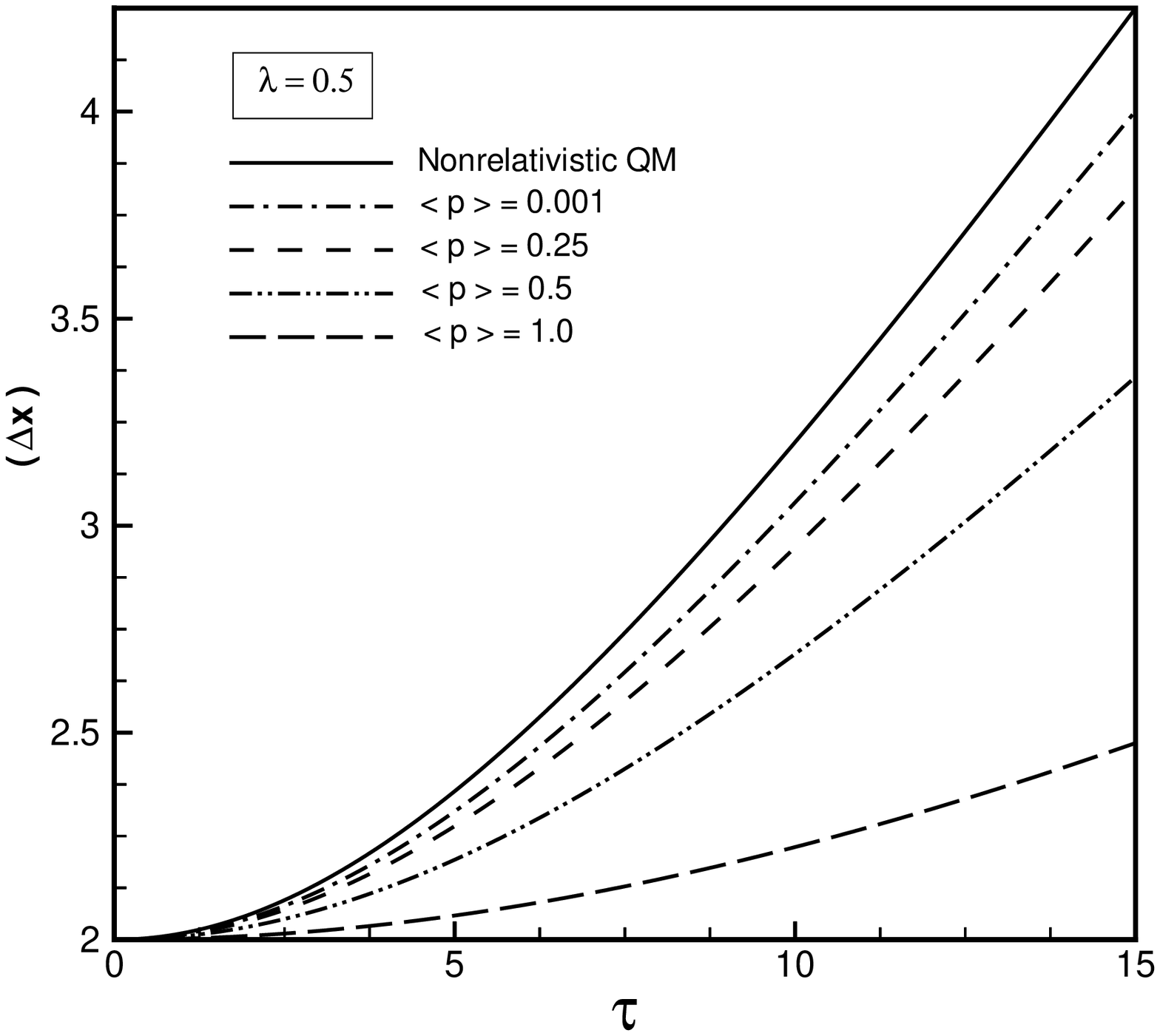}
\hspace{-1.0cm}
\includegraphics[width=.5\columnwidth]{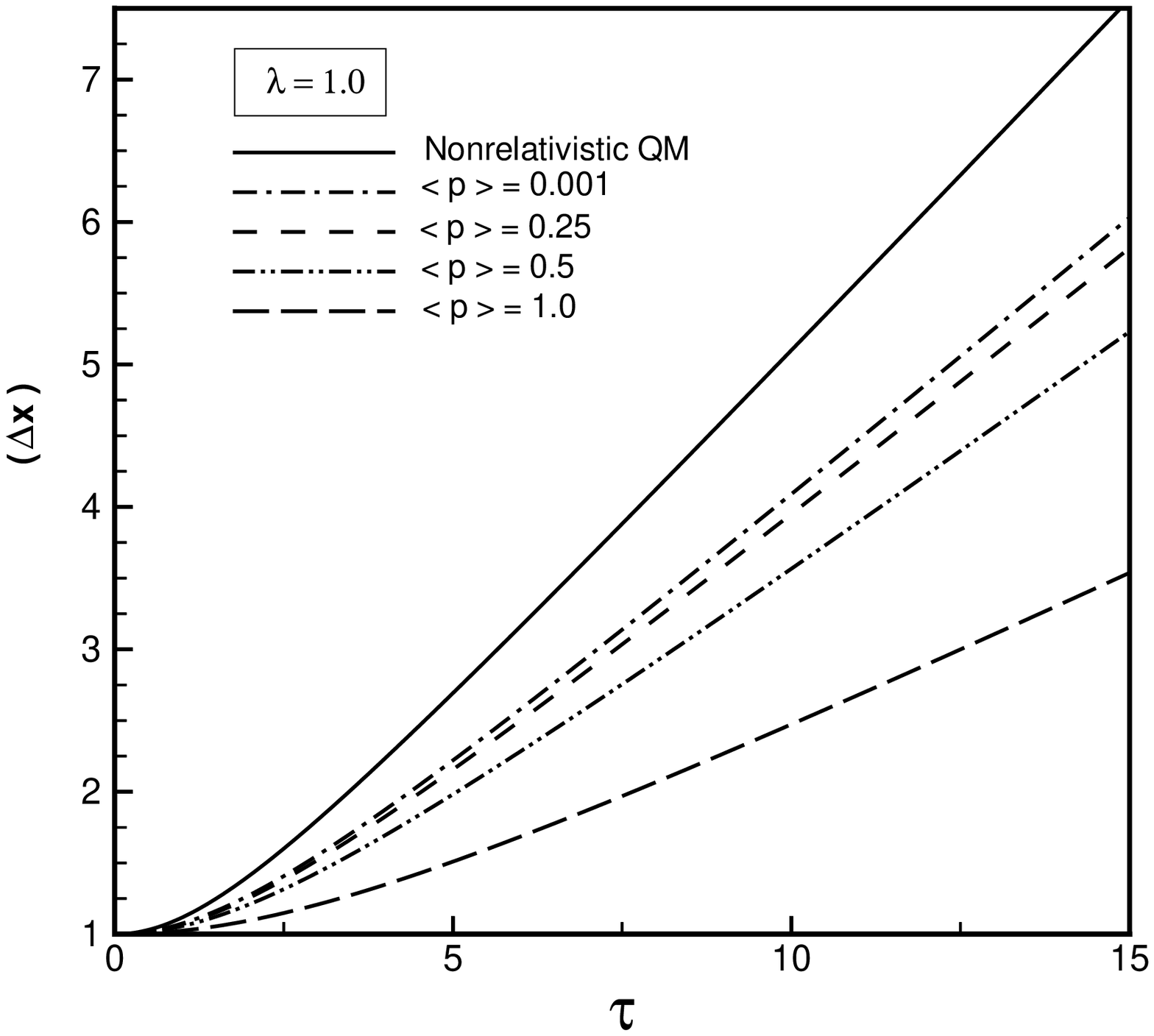}}
\centerline{\includegraphics[width=.5\columnwidth]{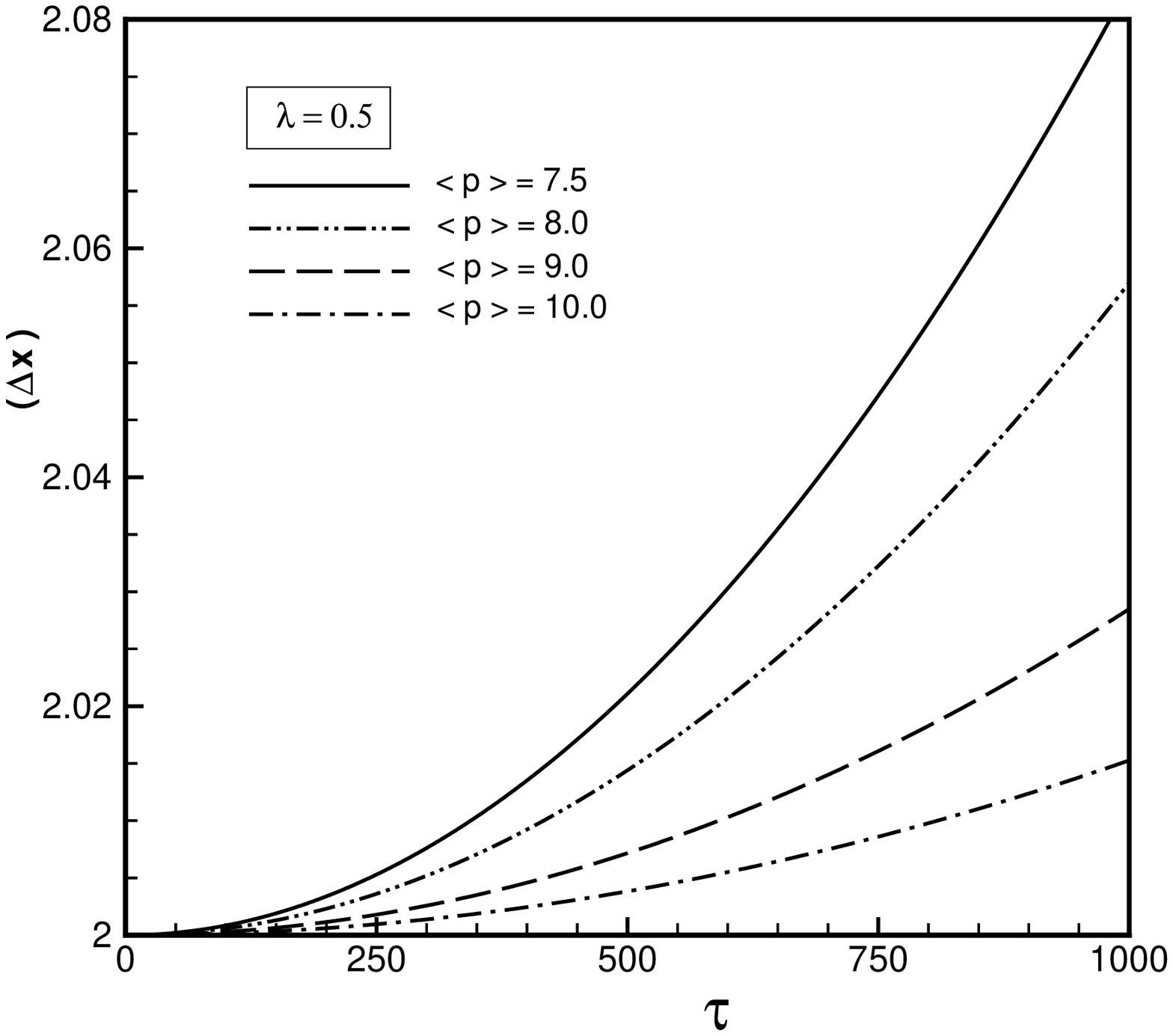}
\hspace{-1.0cm}
\includegraphics[width=.5\columnwidth]{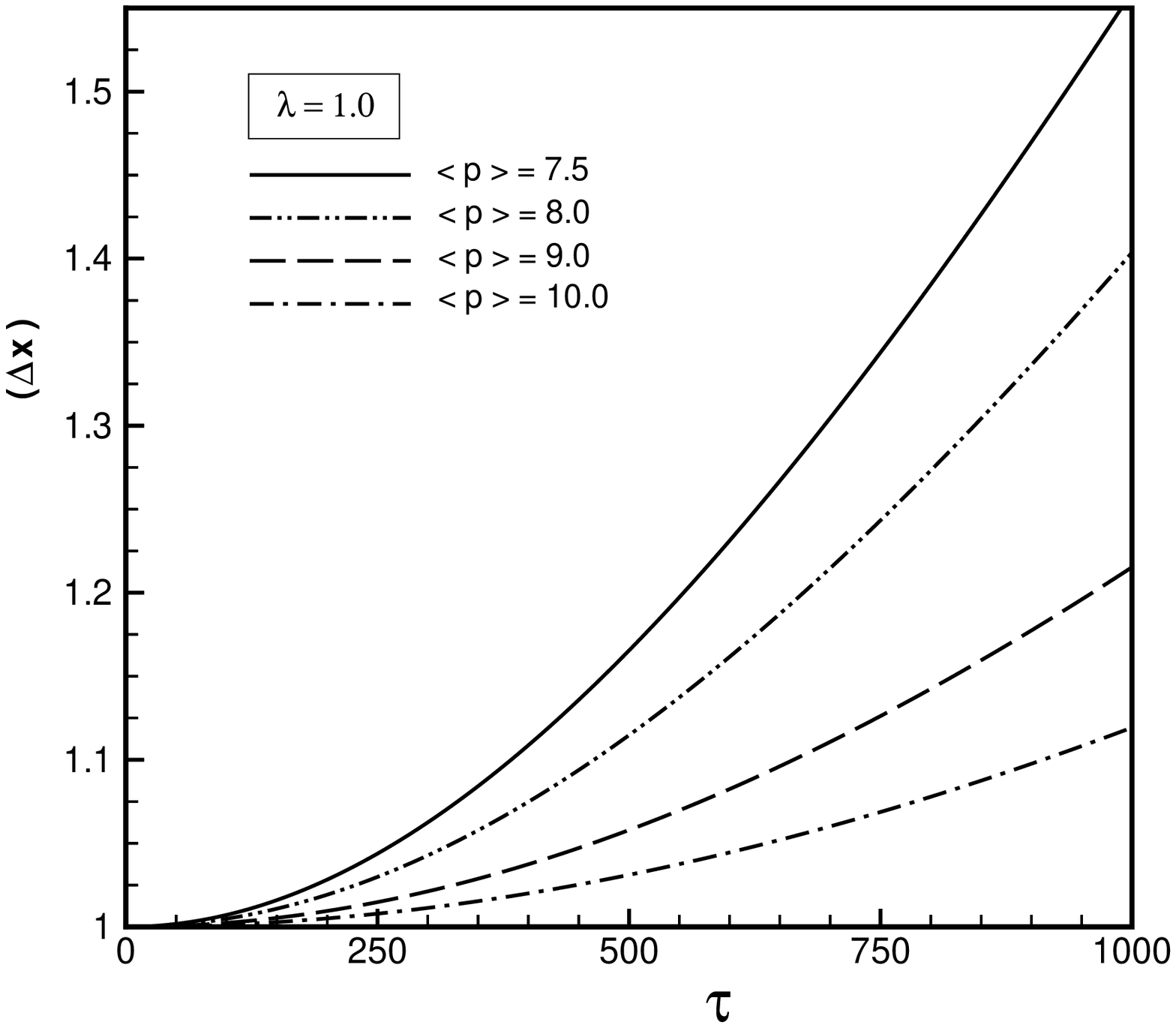}}
\caption{Plots of the dispersion in position $\Delta {\rm x}$ as a
function of time for various $\lambda$ and momentum expectation
values $\br \rp\kt$: The nonrelativistic dispersion which turns
out to be larger than the relativistic dispersion is also
depicted. The higher $\br \rp\kt$ becomes the smaller $\Delta {\rm
x}$ is.} \label{dxfig}
\end{figure}

\begin{figure}
\centerline{\includegraphics[width=.5\columnwidth]{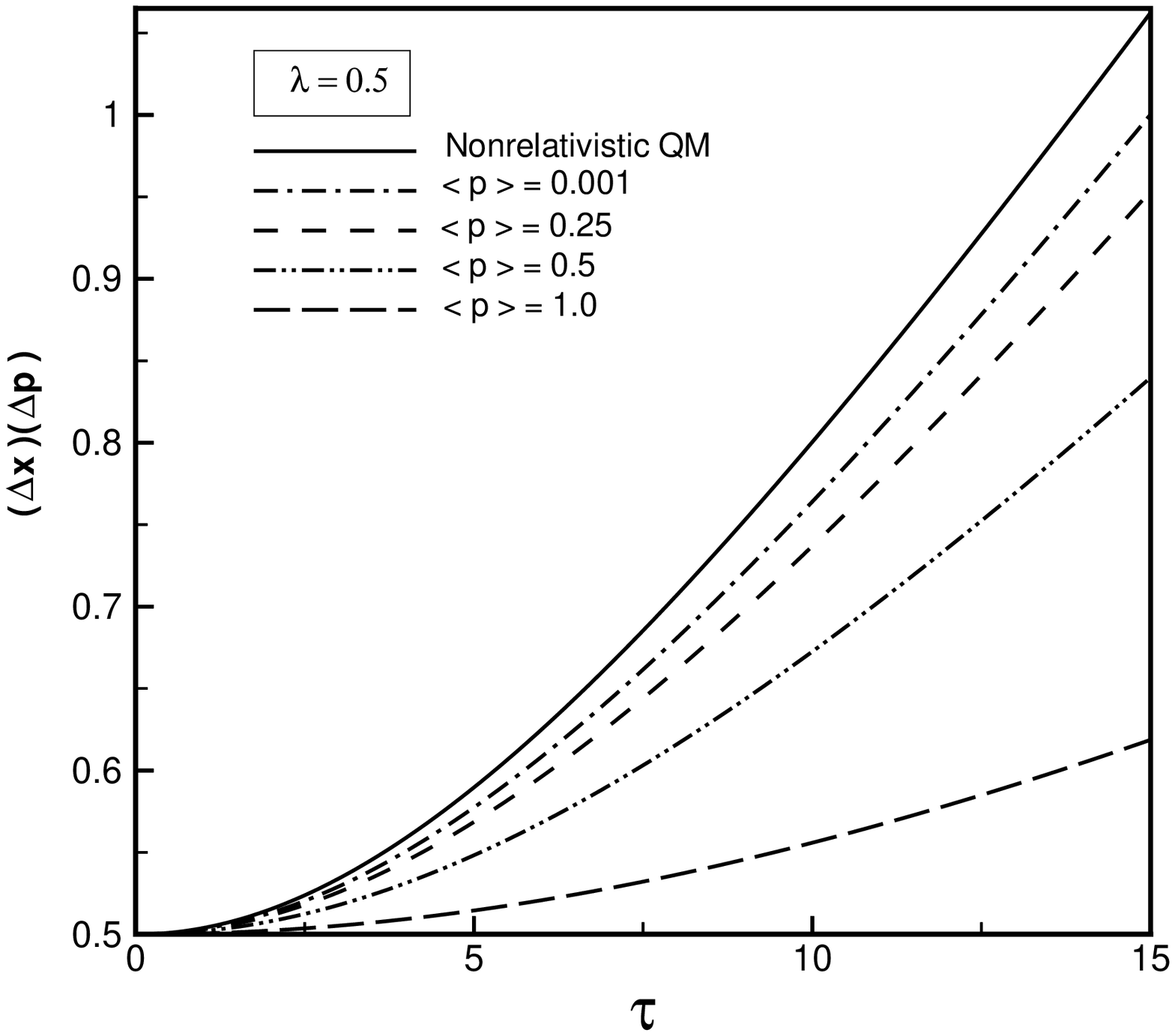}
\hspace{-1.0cm}
\includegraphics[width=.5\columnwidth]{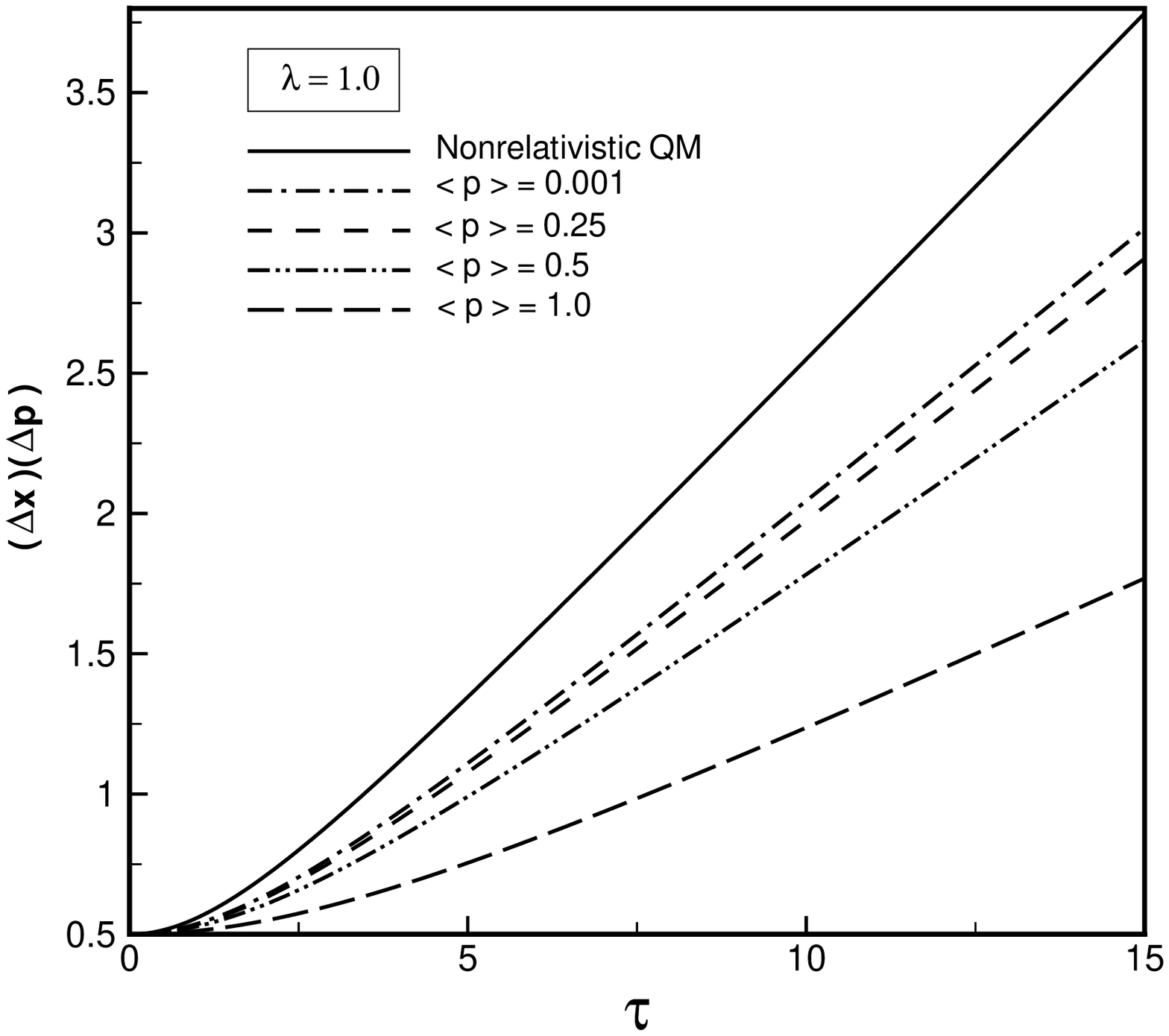}}
\centerline{\includegraphics[width=.5\columnwidth]{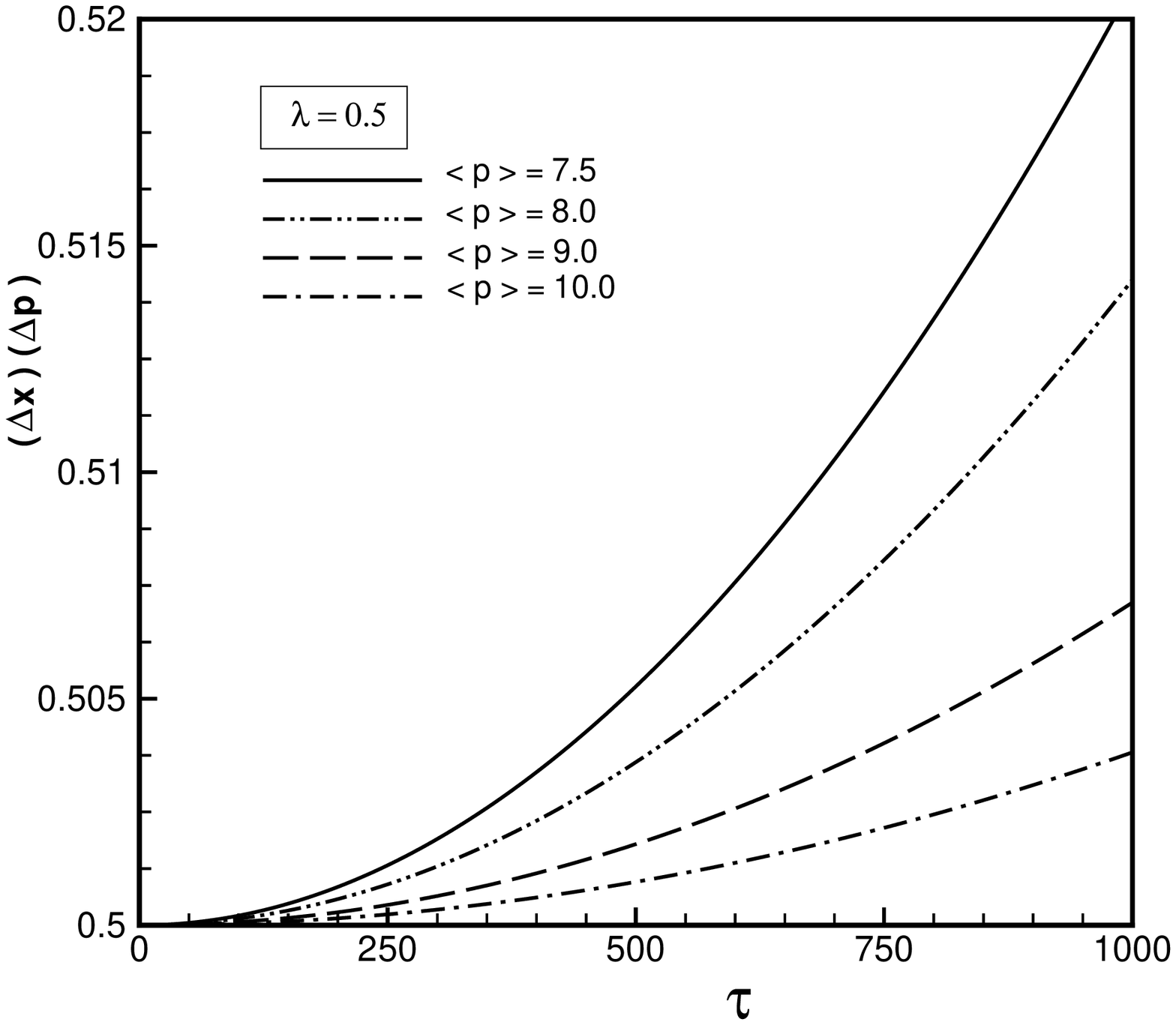}
\hspace{-1.0cm}
\includegraphics[width=.5\columnwidth]{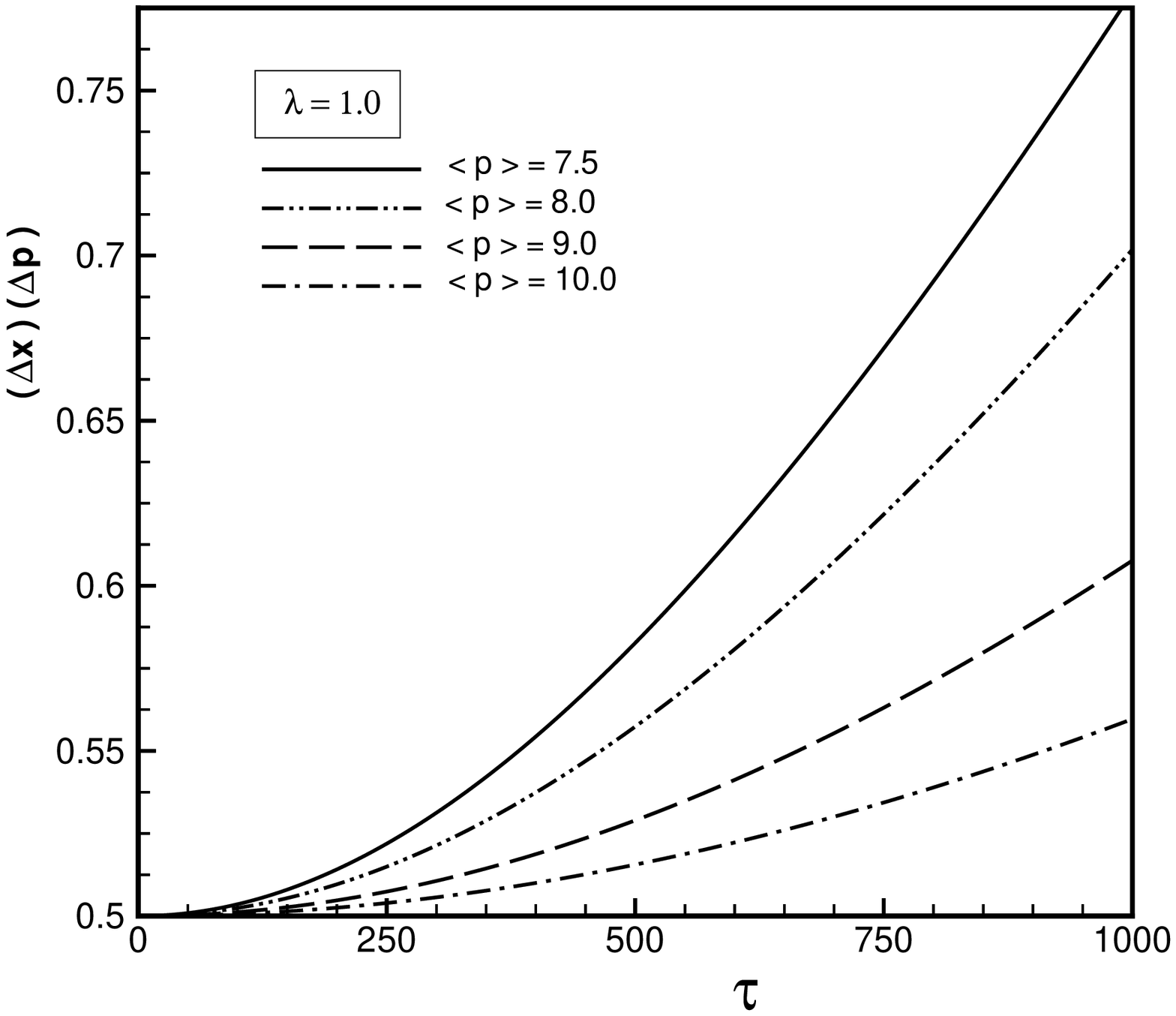}}
\caption{Plots of $(\Delta {\rm x})(\Delta {\rm p})$ in terms of
time for various $\lambda$ and momentum expectation values $\br
\rp\kt$: Nonrelativistic QM yields a larger value for $(\Delta
{\rm x})(\Delta {\rm p})$ than the relativistic QM. The higher the
momentum becomes the smaller $(\Delta {\rm x})(\Delta {\rm p})$
gets. The packets with smaller width (larger $\lambda$) yield
larger values for $(\Delta {\rm x})(\Delta {\rm p})$.}
\label{dxdpfig}
\end{figure}

\begin{figure}
\centerline{\includegraphics[width=.5\columnwidth]{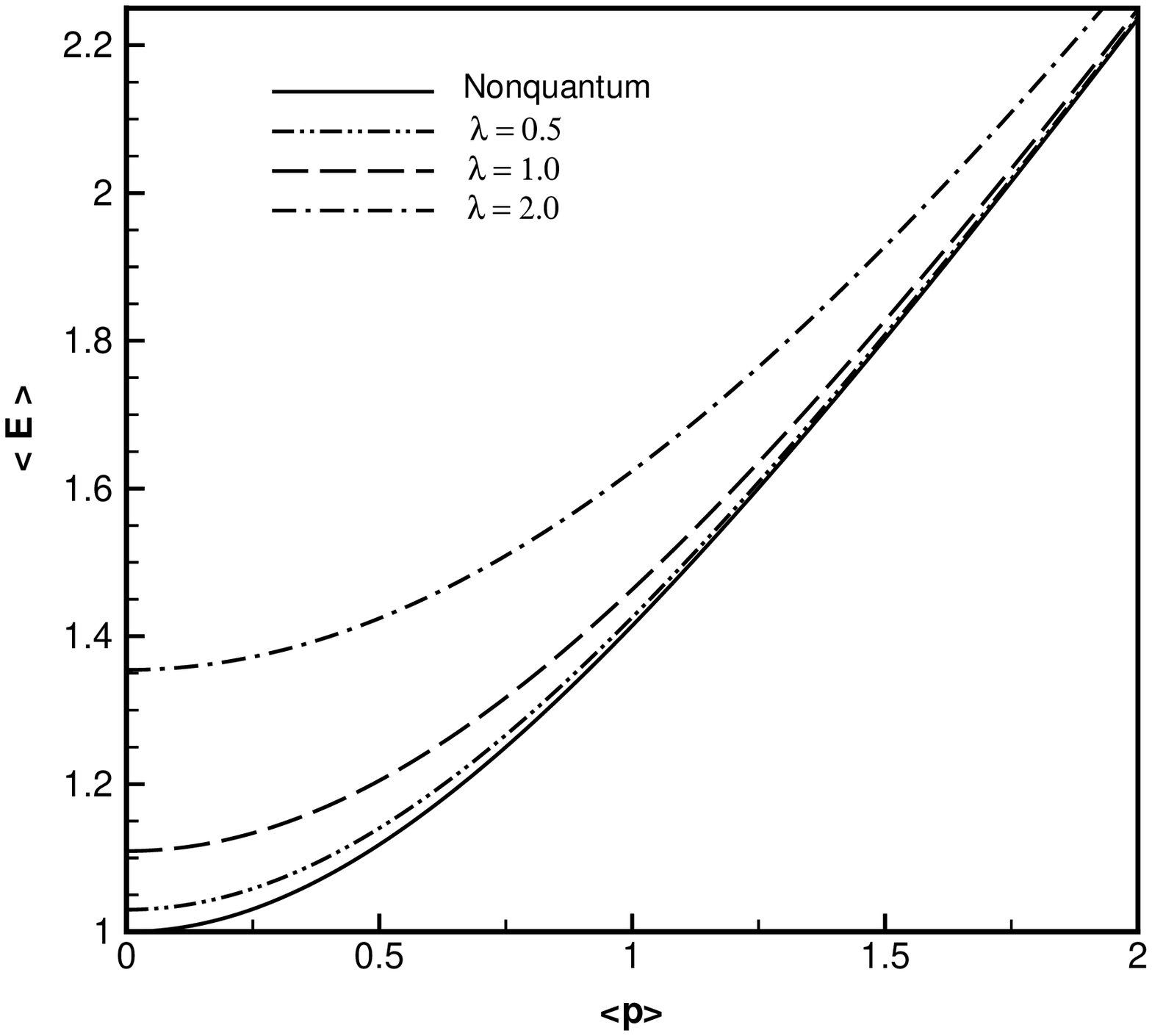}
\hspace{-1.0cm}
\includegraphics[width=.5\columnwidth]{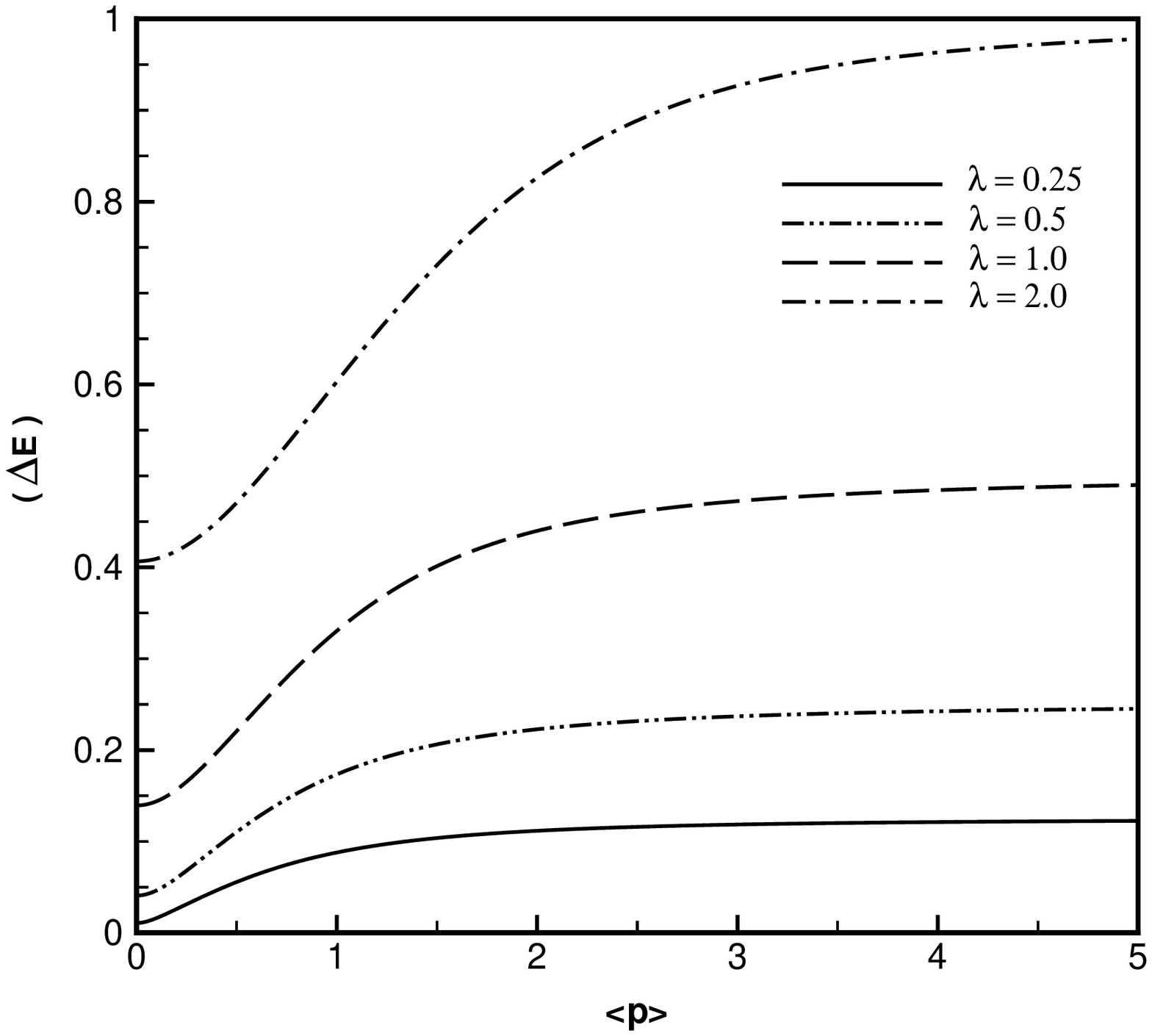}}
\caption{Graphs of the energy expectation value $\br E \kt$ (left)
and its dispersion $(\Delta E)$ (right) as functions of the
momentum expectation value $\br{\rm p}\kt$ for the coherent states
of a free particle with different $\lambda$: The graphs of the
corresponding classical (nonquantum) and nonrelativistic (quantum)
curves are also given. For small values of $\lambda$ the coherent
state displays completely classical behavior.} \label{energy}
\end{figure}

\begin{figure}
\centerline{\includegraphics[width=.95\columnwidth]{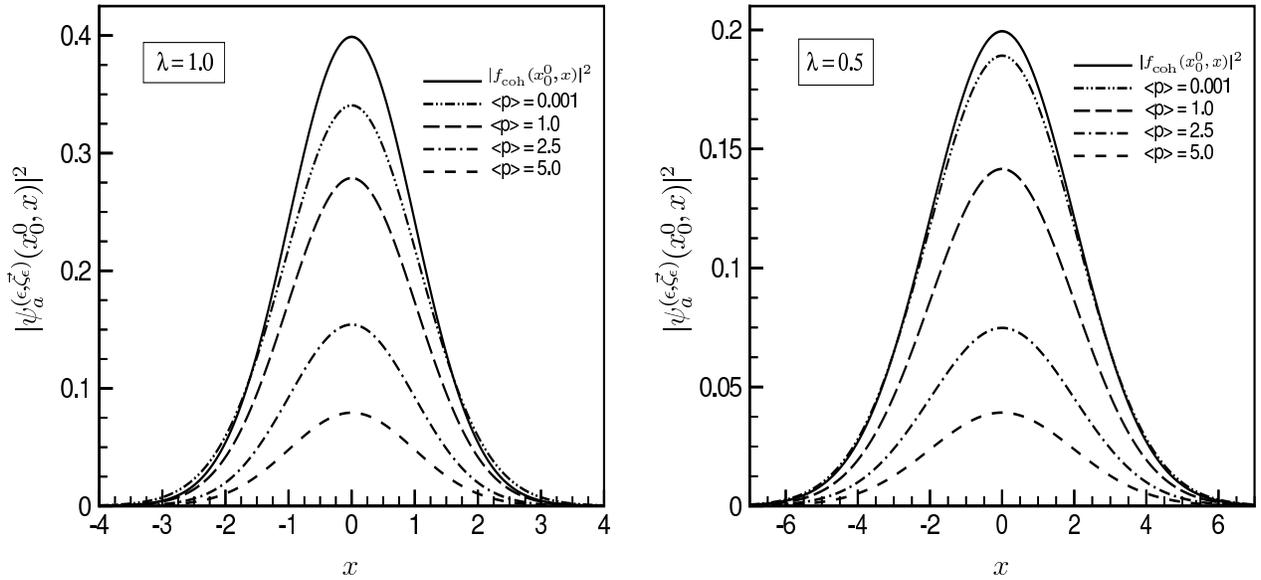}}
\caption{Plots of $|\fc(\epsilon; x^0_0,x)|^2$ and
$|\psiac(x^0_0,x)|^2$ for various $\lambda$ and momentum
expectation values $\br{\rm p}\kt$: $|\fc(\epsilon;x^0_0,x)|^2$
represents the nonrelativistic probability density. For small
values of the momentum expectation value $|\psiac(x^0_0,x)|^2$
tends to $|\fc(\epsilon;x^0_0,x)|^2$. Here we have set
$\kappa(1+\epsilon a)=1$.} \label{probt0}
\end{figure}

\begin{figure}
\centerline{\includegraphics[width=.95\columnwidth]{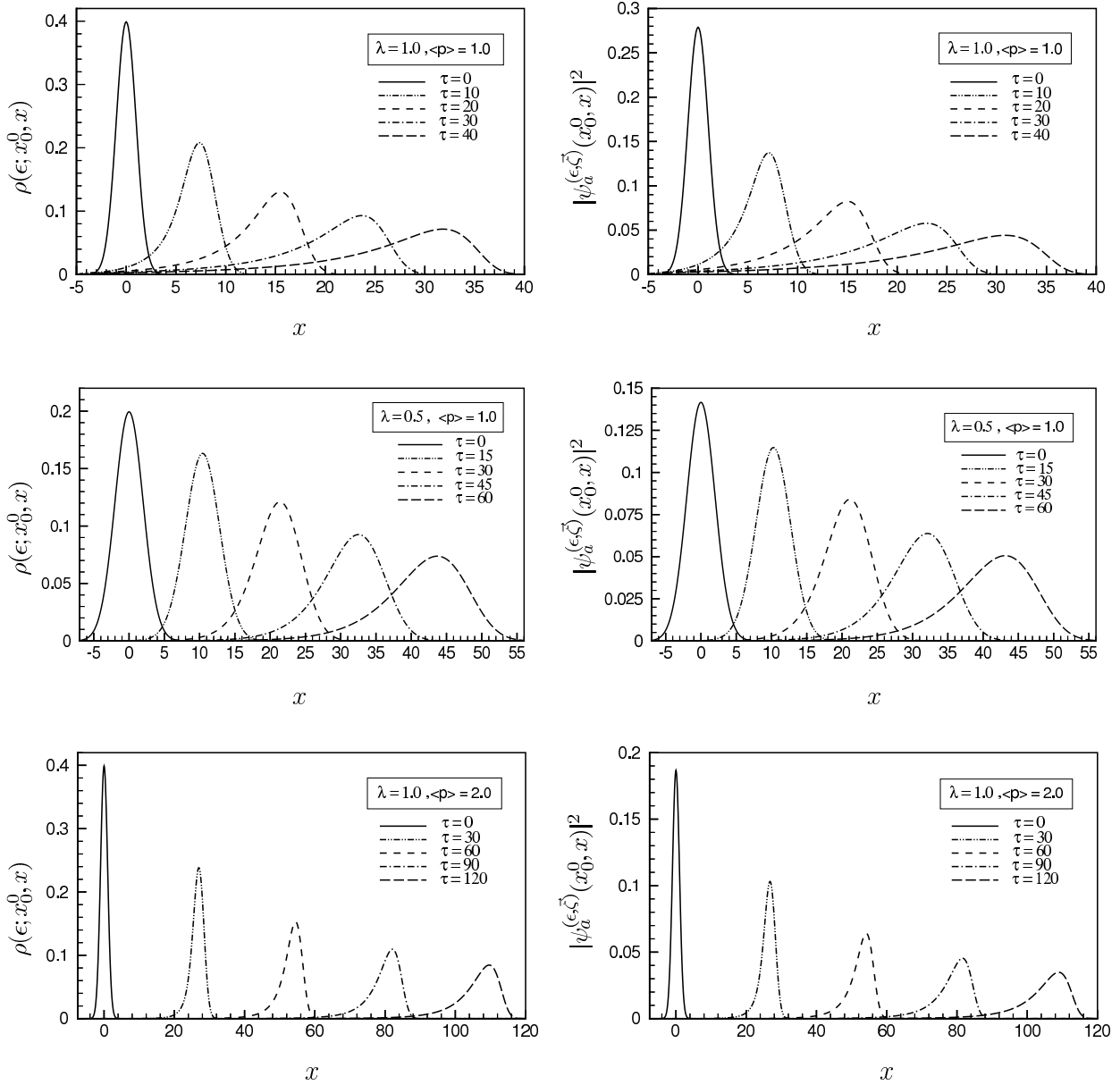}}
\caption{{Graphs showing the time-evolution of the probability
density $\rho(\epsilon; x^0,x)$ and $|\psiac(x^0,x)|^2$ for
various values of $\lambda$, initial $\br\rp\kt$, and
$\epsilon=+1$: They spread with time $\tau$ while their maximum
value decreases. The packets with smaller width spread faster
confirming the behavior displayed in Figs.~(\ref{dxfig}) and
(\ref{dxdpfig}). The faster moving packets behave more like a
classical particle. Unlike for a nonrelativistic coherent wave
packet, the maximum of the probability density does not move with
either of the velocity of the corresponding classical particle or
the velocity expectation value of the packet. It moves with a
higher velocity which is nevertheless smaller than $c$. For a
quantitative analysis see footnote~\ref{foot7}. Here we have set
$\kappa(1+\epsilon a)=1$.}} \label{probtp1}
\end{figure}

\begin{figure}
\centerline{\includegraphics[width=.95\columnwidth]{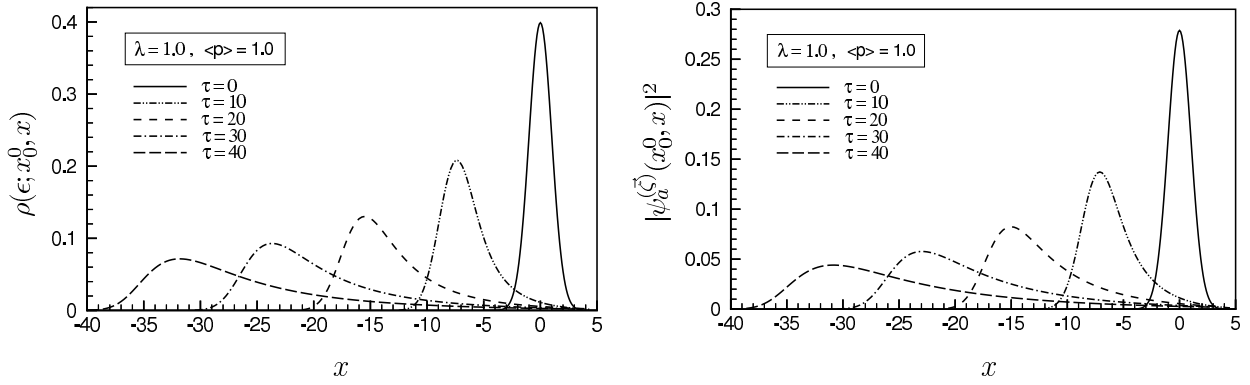}}
\caption{Graphs showing the time-evolution of the probability
density $|\rho(\epsilon; x^0,x)|^2$ and $|\psiac(x^0,x)|^2$ for
$\epsilon=-1$ and a positive initial momentum $\br{\rm p}\kt=1.0$:
The probability density and the coherent KG wave function evolve
in $-x^0$ direction. Here we have set $\kappa(1+\epsilon a)=1$.}
\label{probepsm}
\end{figure}

\begin{figure}
\centerline{
\includegraphics[width=.85\columnwidth]{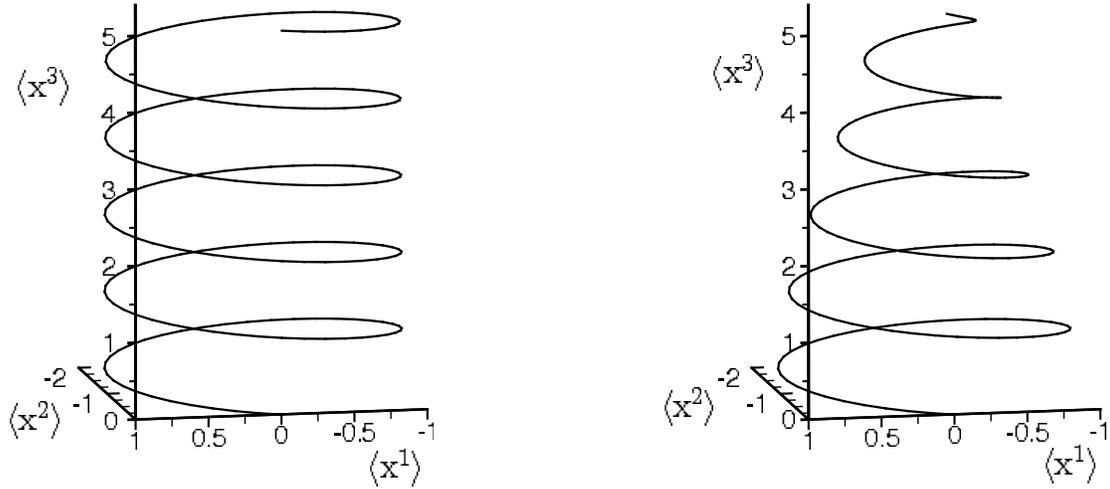}}
\caption{Typical trajectories traced by the expectation value of
the position operator for a coherent state in a magnetic field
with magnetic field parameter $\Lambda = 0.001$, $\br\Pi\kt = 2,
\br{\rm p}_3\kt = 1.6$,
$\lambda_3=10^{-3},\lambda_\perp\simeq\sqrt{\Lambda}$ (left) and
$\lambda_3=\lambda_\perp=0.25$ (right). $\br {\rm x}^1\kt$ and
$\br {\rm x}^2\kt$ are given in unit of classical radius ($R_{\rm
cl.}=\br{\rm p}_1\kt/\Lambda$), and $\br {\rm x}^3\kt$ is given in
unit of ${\dot x}_3 \tau_{\rm cl.}$, where ${\dot x}_3$ is the
classical velocity in the parallel direction and $\tau_{\rm
cl.}:=2\pi / \omega_B$ is the classical period of
precession.}\label{helix}
\end{figure}

\begin{figure}
\vspace{-1.25cm}
\centerline{\includegraphics[width=.88\columnwidth]{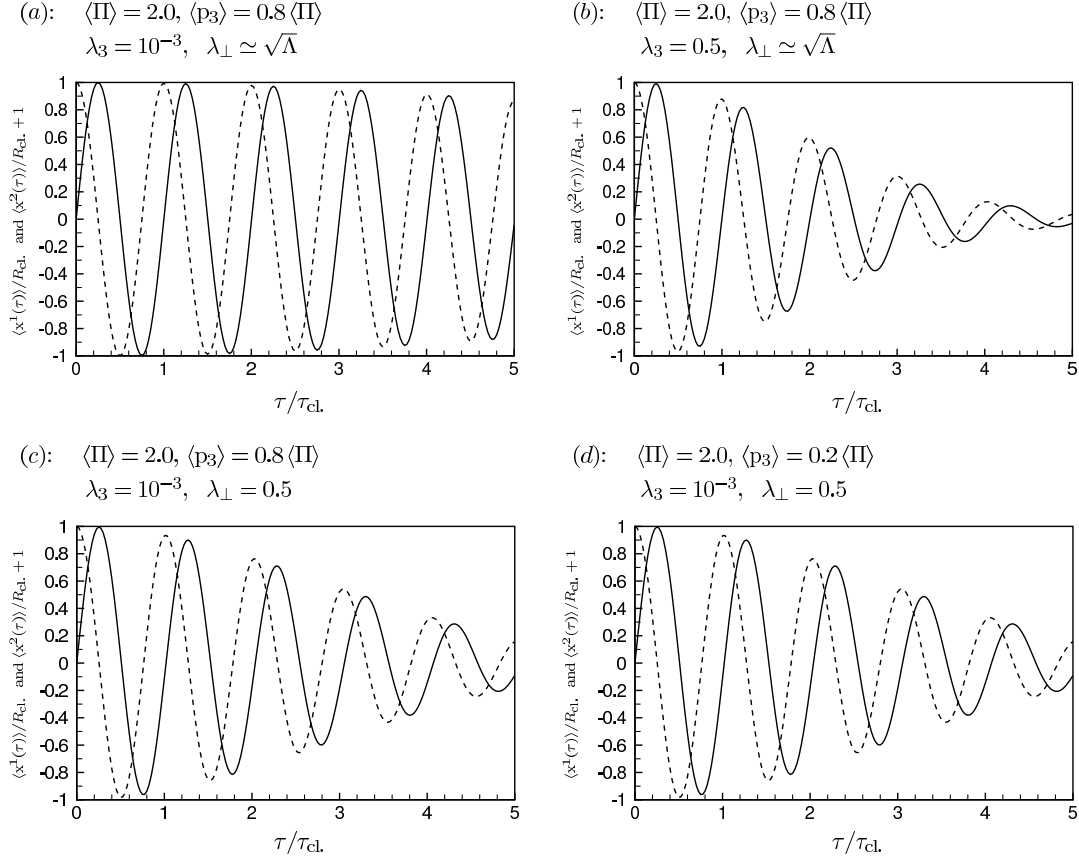}}
\vspace{-0.25cm}
\caption{Plots of $\br{\rm x}^1(\tau)\kt$ (solid
curve) and $\br{\rm x}^2(\tau)\kt+1$ (dashed curve) for various
widths $\lambda_\perp$ and $\lambda_3$ with $\Lambda=0.01$, and
$\br\Pi\kt=2.0$. $\br{\rm x}^1(\tau)\kt$ and $\br{\rm
x}^2(\tau)\kt$ are scaled with the classical radius $R_{\rm cl.}$
and oscillate, with a decreasing amplitude, respectively like
damped sine and cosine functions of the scaled time
$\tau/\tau_{\rm cl.}$. $\tau_{\rm cl.}$ is classical period of
precession. $\lambda_\perp\simeq\sqrt{\Lambda}$ means
$|\lambda_\perp - \sqrt{\Lambda}| \leq 10^{-4}$. For
$\lambda_\perp\rightarrow\sqrt{\Lambda}$ and small value of
$\lambda_3$ the curves approach to the corresponding classical
curves. The effect of $\lambda_3$ dominates that of
$\lambda_\perp$. Also changing the initial transverse momentum
does not affect the behavior of $\br{\rm x}^1(\tau)\kt$ and
$\br{\rm x}^2(\tau)\kt$, though their magnitude clearly depends on
the initial transverse momentum.} \label{x1x2-1}
\end{figure}
\begin{figure}
\vspace{-0.25cm}
\centerline{\includegraphics[width=.88\columnwidth]{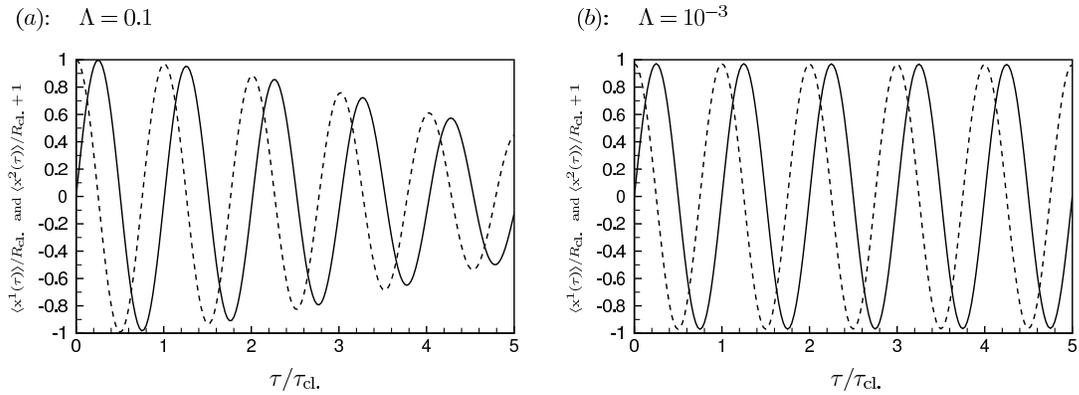}}
\caption{Plots of $\br{\rm x}^1(\tau)\kt$ (solid curve) and
$\br{\rm x}^2(\tau)\kt+1$ (dashed curve) for two values of the
magnetic field parameter $\Lambda=0.1, 0.001$ with the same
initial momentum $\br\Pi\kt=3, \br{\rm p}_3\kt=2.4$, and widths
$\lambda_3=10^{-3}, \lambda_\perp\simeq\sqrt{\Lambda}$. For
$\lambda_\perp\rightarrow\sqrt{\Lambda}$ and small value of
$\lambda_3$ the curves traced by the expectation value of the
position operator approach the classical curves.} \label{x1x2-2}
\end{figure}
\begin{figure}
\vspace{-1.0cm}
\centerline{\includegraphics[width=.95\columnwidth]{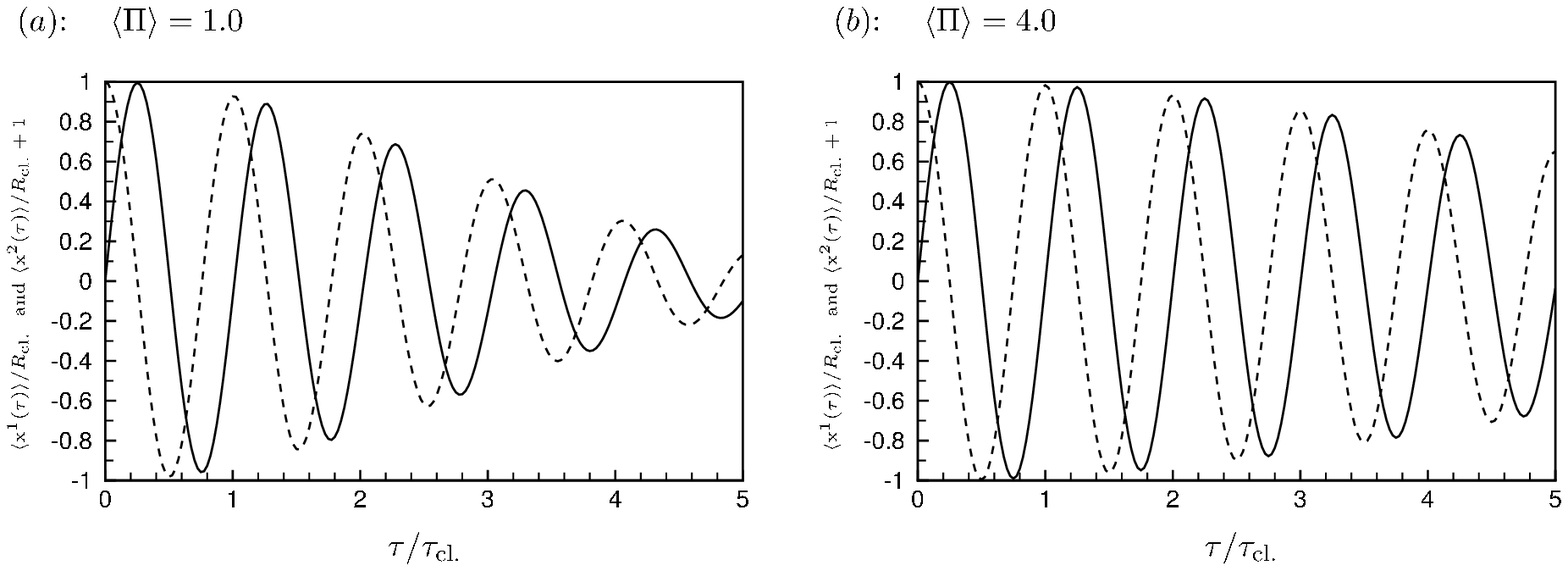}}
\caption{Plots of $\br{\rm x}^1(\tau)\kt$ (solid curve) and
$\br{\rm x}^2(\tau)\kt+1$ (dashed curve) for $\Lambda=0.01$,
$\br\Pi\kt=1$~and~$4$, $\lambda_3=\lambda_\perp=0.25$ and $\br{\rm
p}_3\kt=0.8\br\Pi\kt$. $\br{\rm x}^1(\tau)\kt$ and $\br{\rm
x}^2(\tau)\kt$ are scaled with the classical radius and time is
scaled with the classical period of precession. The higher the
initial momentum becomes the closer the curves get to the
corresponding classical curves.} \label{x1x2-3}
\end{figure}
\begin{figure}
\vspace{-2.0cm}
\centerline{\includegraphics[width=.95\columnwidth]{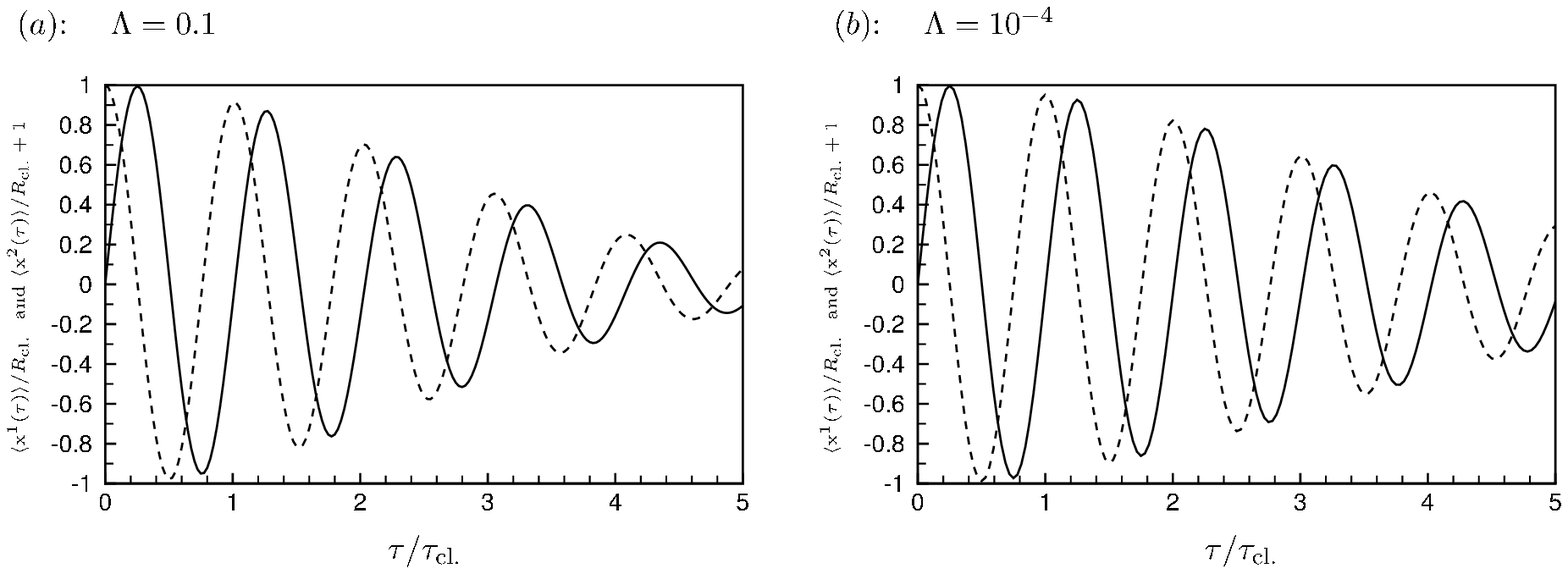}}
\caption{Plots of $\br{\rm x}^1(\tau)\kt$ (solid curve) and
$\br{\rm x}^2(\tau)\kt+1$ (dashed curve) for
$\Lambda=0.1$~and~$10^{-4}$, $\br\Pi\kt=2, \br{\rm p}_3\kt=1.6$,
and $\lambda_3=\lambda_\perp=0.25$. $\br{\rm x}^1(\tau)\kt$ and
$\br{\rm x}^2(\tau)\kt$ are scaled with the classical radius and
time is scaled with the classical period of precession. As the
magnetic field decreases the curves approach the corresponding
classical curves.} \label{x1x2-4}
\end{figure}
\begin{figure}
\centerline{\includegraphics[width=.95\columnwidth]{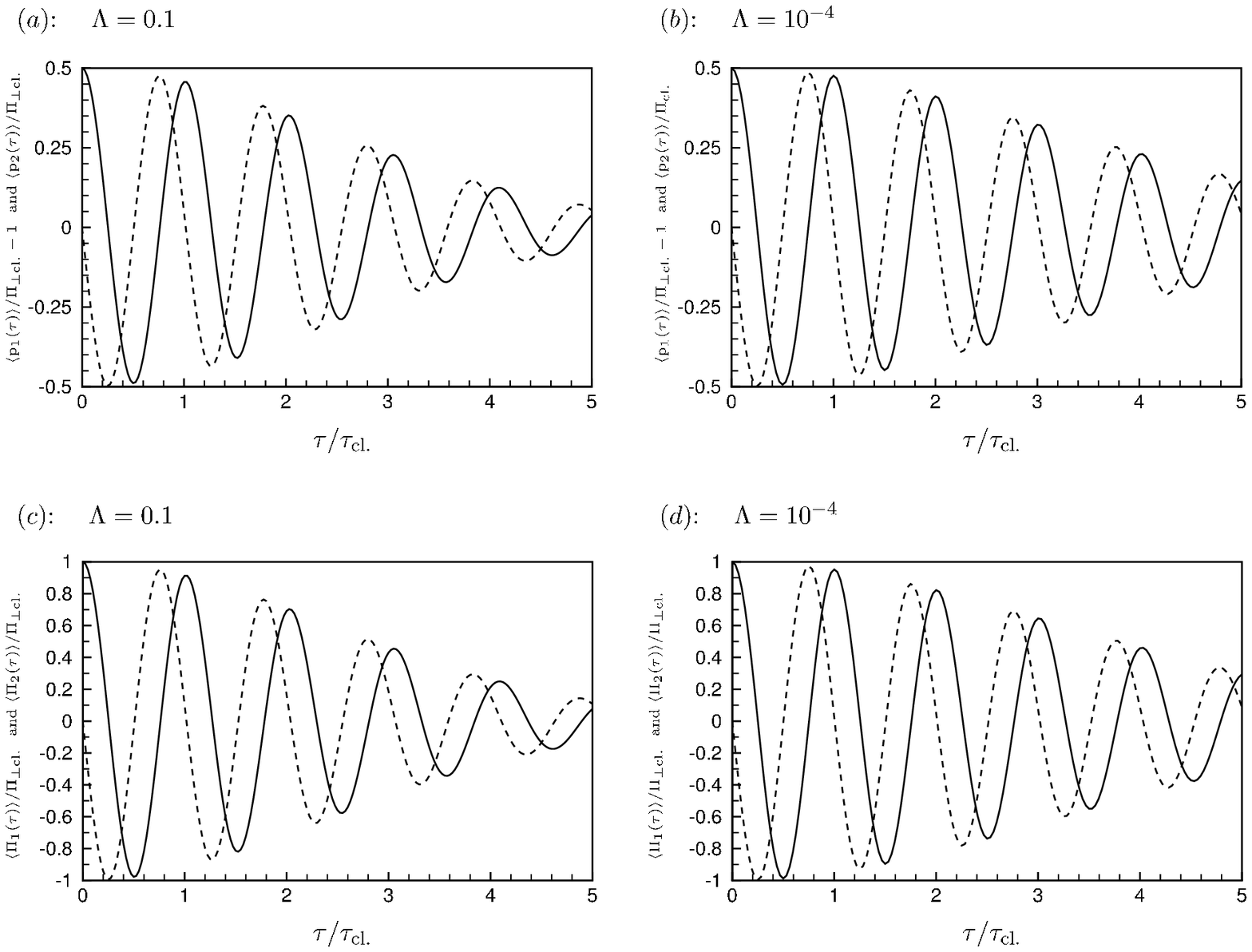}}
\caption{($a$) and ($b$) are plots of $\br{\rm p}_1(\tau)\kt-1$
(solid curve), $\br{\rm p}_2(\tau)\kt$ (dashed curve), and ($c$)
and ($d$) are plots of the transverse kinetic momentums
$\br\Pi_1(\tau)\kt$ (solid curve) and $\br\Pi_2(\tau)\kt$ (dashed
curve), for $\Lambda=0.1$~and~$10^{-4}$, $\br\Pi\kt=2, \br{\rm
p}_3\kt=1.6$, and $\lambda_3=\lambda_\perp=0.25$. The expectation
value of the momentum operators are scaled with the corresponding
classical transverse kinetic momentum $\Pi_\perp=\br{\rm p}_1\kt$
and the time is scaled with the classical period of precession.
The behavior of $\br{\rm p}_1(\tau)\kt-1$ and $\br\Pi_1(\tau)\kt$
(resp.\ $\br{\rm p}_2(\tau)\kt$, and $\br\Pi_2(\tau)\kt$)
reminisce the damped cosine (resp.\ sine) function. According to
Eq.~(\ref{f-mat-p123}) the behavior of $\br{\rm p}_i\kt$ (and
consequently $\br\Pi_i\kt$) is similar to that of $\br{\rm
x}_i\kt$. Therefore, the higher the initial momentum (and the
smaller the magnetic field) is the closer the curves traced by the
expectation value of momentum operators are to the corresponding
classical curves. Also for
$\lambda_\perp\rightarrow\sqrt{\Lambda}$ and small value of
$\lambda_3$ the curves approach to the classical ones.}
\label{p1p2}
\end{figure}
\begin{figure}
\centerline{\includegraphics[width=.95\columnwidth]{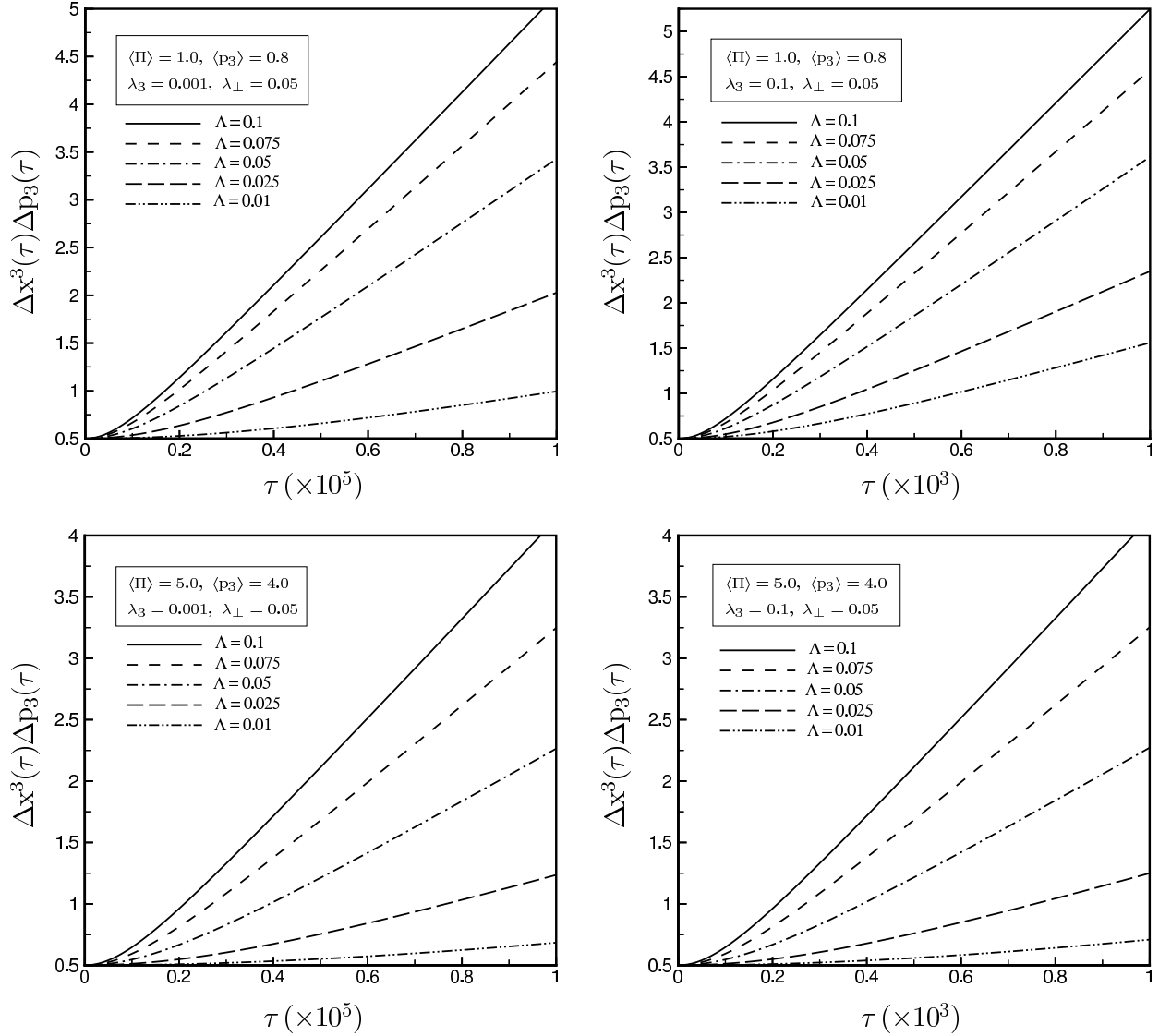}}
\caption{Plots of $(\Delta {\rm x}^3)(\Delta {\rm p}_3)$ as a
function of time for different initial momentum expectation
values, magnetic field parameter $\Lambda$, and the widths
$\lambda_3$ and $\lambda_\perp$. $(\Delta {\rm x}^3(\tau))(\Delta
{\rm p}_3(\tau))$ is an increasing function of the magnetic field.
The presence of the magnetic field enhances the spreading of the
wave packet. Also faster moving wave packets have a slower
spreading rate.} \label{magunc}
\end{figure}

\begin{figure}
\vspace{-0.5cm} \centerline{
\includegraphics[width=.55\columnwidth]{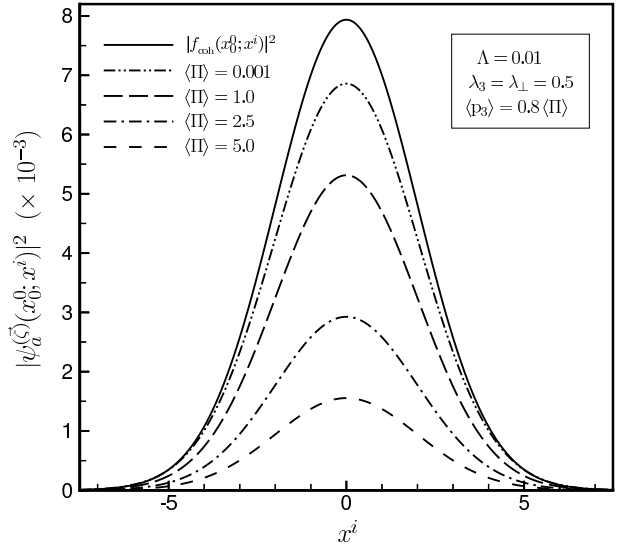}}
\caption{Plots of $|\psi_a^{(\vec\zeta)}(x^0_0;x^i)|^2$ as a
function of $x^i$ for $\Lambda=0.01$, where
$|\psi_a^{(\vec\zeta)}(x^0_0;x^i)|^2$ means
$|\psi_a^{(\vec\zeta)}(x^0_0;\vec x)|^2$ with $x^j=0$ if $j\neq
i$. We has set $\kappa(1+\epsilon a)=1$. Since we consider the
same widths along all three dimensions, (graphs of)
$|\psi_a^{(\vec\zeta)}(x^0_0;x^i)|^2$ are identical for all
$i=1,2,3$. $|\fc(x^0_0;x^i)|^2$, which is shown by the solid line,
gives the nonrelativistic probability density. For smaller values
of the momentum expectation value, i.e., in the nonrelativistic
limit, $|\psi_a^{(\vec\zeta)}(x^0_0;x^i)|^2$ tends to
$|\fc(x^0_0;x^i)|^2$.} \label{kgx3rho-pvar}
\end{figure}
\begin{figure}
\vspace{-0.5cm}
\centerline{\includegraphics[width=.9\columnwidth]{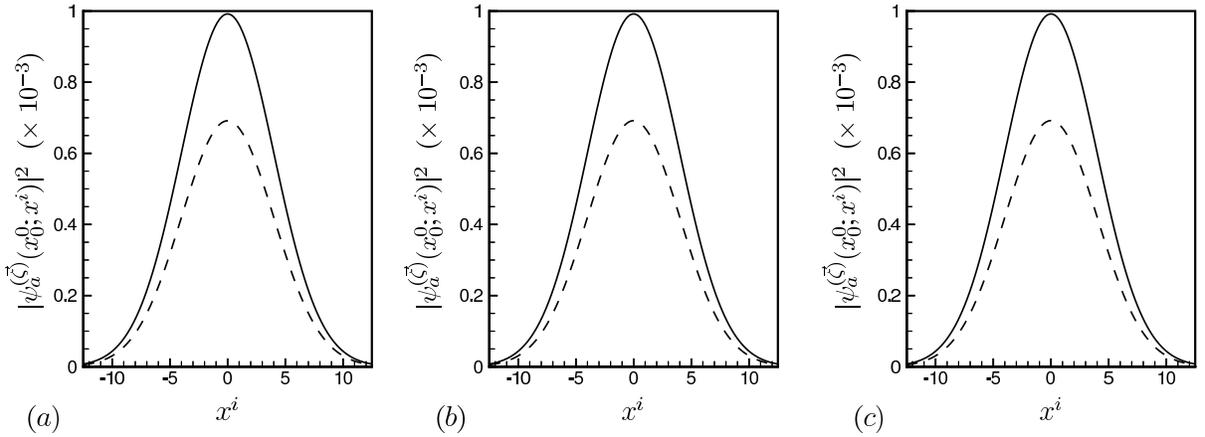}}
\caption{Plots of $|\psi_a^{(\vec\zeta)}(x^0_0;x^i)|^2$ (dashed
curve) with $\br\Pi\kt=1$ and $\lambda_\perp=\lambda_3=0.25$, for
three values of the magnetic field parameter: $\Lambda=0.1$ ($a$),
$\Lambda=10^{-3}$ ($b$) and $\Lambda=10^{-6}$ ($c$). The solid
curve shows the nonrelativistic counterpart $|\fc(x^0_0;x^i)|^2$.
$|\psi_a^{(\vec\zeta)}(x^0_0;x^i)|^2$ does not depend on the
magnetic field parameter.} \label{kgx3rho-mvar}
\end{figure}

\begin{figure}
\centerline{\includegraphics[width=0.95\columnwidth]{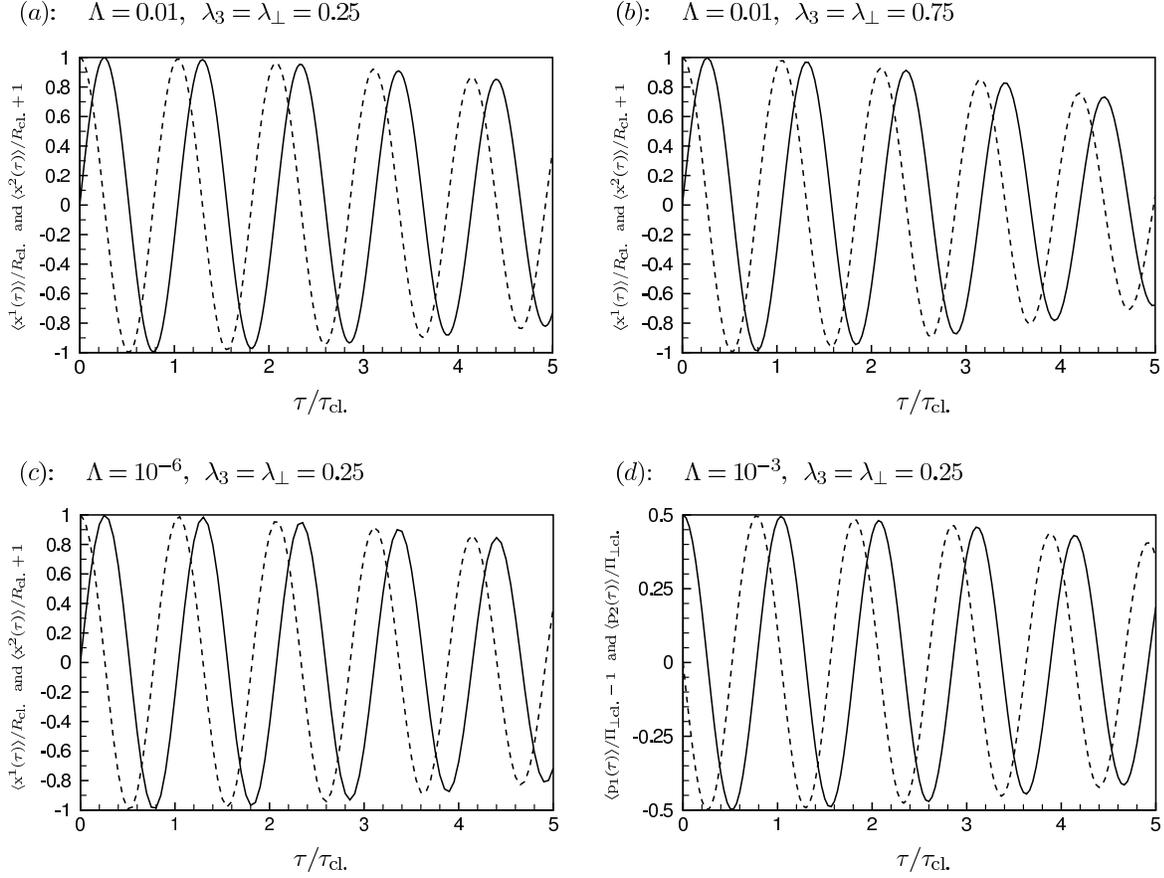}}
\caption{($a$) -- ($c$) are plots of $\br{\rm x}^1(\tau)\kt$
(solid curve) and $\br{\rm x}^2(\tau)\kt+1$ (dashed curve); ($d$)
is the plots of $\br{\rm p}_1(\tau)\kt-1$ (solid curve) and
$\br{\rm p}_2(\tau)\kt$ (dashed curve), for $\br\Pi\kt=10^{-3}$,
different values of the magnetic field parameter $\Lambda$, and
widths $\lambda_3, \lambda_\perp$. The expectation value of
position and momentum operators are scaled respectively with the
classical radius and classical transverse kinetic momentum. Time
is scaled with the classical period of precession. In the
nonrelativistic limit ($c\rightarrow\infty$), the curves do not
depend on either of the widths $\lambda_\perp$ and $\lambda_3$ or
the magnetic field parameter $\Lambda$. They tend to the
corresponding nonrelativistic curves obtained from
Eqs.~(\ref{x123-nr}) and (\ref{p123-nr}) which agree with the
predictions of the classical theory.} \label{x1x2-nr}
\end{figure}

\ed